\PassOptionsToPackage{usenames,dvipsnames}{xcolour}
\documentclass[twocolumn,twocolappendix,nofootinbib,iop,numberedappendix]{openjournal}
\usepackage{graphicx}
\usepackage{amsmath,amssymb,amstext}
\usepackage[utf8]{inputenc}
\usepackage[colorlinks,allcolors=blue]{hyperref}
\usepackage{natbib}
\usepackage{url}
\usepackage{lineno}
\usepackage{xcolor}
\usepackage{apjfonts}

\begin{document}
\journalinfo{The Open Journal of Astrophysics}

%%%%%%%%%%%%%%%%%%%%%%%%%%%%%%%%%%%%%%%%%%%%%%%%%%%
% Title
%%%%%%%%%%%%%%%%%%%%%%%%%%%%%%%%%%%%%%%%%%%%%%%%%%%
\title{\vspace{-0.5cm}Demographics and growth of the astronomy and astrophysics community\vspace{-1.4cm}}

%%%%%%%%%%%%%%%%%%%%%%%%%%%%%%%%%%%%%%%%%%%%%%%%%%%
% Author list
%%%%%%%%%%%%%%%%%%%%%%%%%%%%%%%%%%%%%%%%%%%%%%%%%%%
\author{J\"urgen Kn\"odlseder$^{1,\ast}$}
\thanks{$^\ast$jurgen.knodlseder@utoulouse.fr}
\affiliation{$^1$Institut de Recherche en Astrophysique et Plan\'etologie, Universit\'e de Toulouse, CNRS, CNES, UPS,
9 avenue Colonel Roche, 31028 Toulouse, Cedex 4, France}

%\date{Received 27 October 2026}
%\date{Received 1 November 2026 / Accepted 1 December 2026}

%%%%%%%%%%%%%%%%%%%%%%%%%%%%%%%%%%%%%%%%%%%%%%%%%%%
% Abstract
%%%%%%%%%%%%%%%%%%%%%%%%%%%%%%%%%%%%%%%%%%%%%%%%%%%
\begin{abstract}
%% Context
%The scientific literature is growing at an exponential rate, making it increasingly difficult for scientists
%and editors to cope with the flow of scientific publications.
%Astronomy is not exempt from this inflation, yet little is known about the demographics and evolution
%of the community that produces this flow.
%% Aims
%This paper aims to fill this gap by studying the demographics and growth of the astronomy and 
%astrophysics community over the past $\sim$200 years.
%% Method
%The study makes use of  bibliographic information provided by the Astrophysics Data System from 
%which annual author counts, author affiliations, paper and page numbers were extracted.
%A name-matching scheme was developed that allowed to cross-identify authors between publications 
%and to gain insights in activity periods and professional age distributions.
%Affiliation information was used to assess the type of host institutions that the authors are working 
%for and to determine the fraction of publishing amateur astronomers.
This paper presents the demographics and growth of the astronomy and astrophysics community 
over the past $\sim$200 years based on data provided by the Astrophysics Data System. 
% Results
The analysis suggests that 59\,940 authors signed at least one paper in 2025, while about 40\% of 
the community did not publish any papers, resulting in an estimated community size of roughly 
100\,000 astronomers in 2025.
About 38\% of the community is hosted in Europe, followed by Asia (29\%), North America (26\%),
South America (4\%), Oceania (2\%) and Africa (1\%).
Large variations exist among the countries in the number of astronomers, astronomers per capita
of country population and astronomers per country gross domestic product (GDP).
Overall, the community is dominated by early career scientists, with 50\% of the authors publishing
papers for seven years or less.
This dominance is readily explained by a steady and exponential growth of the community of 4.3\% 
per year over the last 50 years, implying that the community-size doubled every 16 years.
The growth rate of the community exceeds that of the number of papers or paper pages, implying
that the annual number of papers or pages published per astronomer is on decline.
Over the last decade, the exponential growth of the community has accelerated to 5.1\% per year,
owing in particular to countries like China (13\% per year) and India (10\% per year).
For almost all countries in the world, growth rates in authors, authors per capita and authors per
GDP are positive.
The growth in astronomy is not an exception, as the growth rates of the earth sciences and physics 
communities are even larger.
% Conclusions
If the past tendencies continue, Chinese and Indian astronomers will be first and second largest 
communities in the world by 2029 and 2043, respectively.
\end{abstract}

%\keywords{
%General: History and philosophy of astronomy --
%General: Publications, bibliography --
%General: Sociology of Astronomy
%}

\maketitle
%\nolinenumbers

%%%%%%%%%%%%%%%%%%%%%%%%%%%%%%%%%%%%%%%%%%%%%%%%%%%
% Introduction
%%%%%%%%%%%%%%%%%%%%%%%%%%%%%%%%%%%%%%%%%%%%%%%%%%%
\section{Introduction}

It is well established that the scientific literature grows at an exponential rate, doubling its size every
ten to fifteen years \citep{price1951}.
For the field of astronomy and astrophysics (hereafter simply ``astronomy''), literature doubling times 
as short as eight years have been reported following World War II \citep{abt1981}.
The fact that the astronomical community is even growing faster, which is actually driving this literature
growth \citep{schmidtkaler1982,davoust1987}, is less well appreciated.
This may be explained by the sparseness of literature on the demographics and evolution of the 
astronomy community, and even the rather simple question of how many astronomers are currently 
active in the world has so far no academic answer.

There exist different methods to assess the size of the astronomical community,
including estimates from national community reports
\citep{jarrell1990,vanderkruit1994a,vanderkruit1994b,ahn2019},
membership counts in astronomical societies such as the International Astronomical Union (IAU) 
\citep{jaschek1992}, 
or author counts on scientific publications 
\citep{schmidtkaler1982,davoust1987,jaschek1991,liu1995,abt1998,kurtz2005}.
Each method comes with its own limitations and implies a different definition of the term
``astronomical community''.
National community reports may be incomplete and do not exist for all countries and years.
Furthermore, they may only include specific categories, such as tenured astronomers or astronomers
working in a limited number of institutes.
Membership in astronomical societies depends on access criteria and curation of membership lists
over the years.
For example, IAU membership is known to be subject to important variations among countries and
to include inactive astronomers \citep{vanderkruit1994a,vanderkruit1994b,hearnshaw2007}.
Author counts are probably least biased and make use of the fact that any professional astronomer 
will sooner or later manifest by having her or his name on the author list of a scientific publication.
Yet they imply a relatively wide definition of ``astronomical community'' since they include every 
person publishing in astronomy, including amateurs, engineers, scientists from neighbouring fields of 
research, journalists, editorialists and managers of scientific research \citep{davoust1987}.

The most comprehensive works to date using author counts were published about 35 years ago
\citep{davoust1987,jaschek1991}.
Both works make use of the author indexes of the Astronomy and Astrophysics Abstracts that 
were published between 1969 and 2000 by the Astronomisches Rechen-Institute 
Heidelberg\footnote{\url{https://wwwadd.zah.uni-heidelberg.de/publikationen/aaa/}}.
\cite{davoust1987} found that the number of authors rose from 8\,278 in 1969 to 19\,208 in 1985,
implying a doubling-time in the number of authors of about 15 years.
About 60\% of the authors published only a few papers over a 15-year interval, from which
\cite{davoust1987} concluded that there exists a large population of ``transient astronomers'' 
comprised in particular of students who do not stay in astronomy.
\cite{jaschek1991} complemented the data by some older sources, inferring for the period 
1952--1980 a doubling-time of the number of authors of about ten years.
Developing a dedicated model for the population of astronomers, he infers that about half of 
the astronomers ``are just occasional paper producers'', confirming the finding of \cite{davoust1987}.
Using his model he came to the conclusion that in the year 1985 there existed about 9\,000 
permanent astronomers in the world.

Since these works the introduction of the Astrophysics Data System 
(ADS\footnote{\url{https://ui.adsabs.harvard.edu/}}) 
in 1992 has made a large 
bibliographic collection in astronomy easily available \citep{kurtz2009}, allowing detailed bibliometric
studies that were not possible before \citep[e.g.][]{kurtz2009,henneken2019}.
Given the availability and easy access to the data it is surprising that, at least to my knowledge, ADS 
was never used to study the demographics and evolution of the astronomy community.
\cite{davoust1987} and \cite{jaschek1991} considered data up to 1985, and during the 40 years that
passed since then, many things have happened that potentially may have impacted the evolution of the
community: 
the USSR has collapsed and was dissolved, 
new actors such as China and India have emerged, 
the world population has risen from 4.9 billion to 8.3 billion, 
climate change and environmental degradations increasingly disrupt human societies, 
and the world currently experiences a major geopolitical reconfiguration.
It seems therefore timely to have a new look at the demographics and growth of the astronomy
community using the latest information available.

%%%%%%%%%%%%%%%%%%%%%%%%%%%%%%%%%%%%%%%%%%%%%%%%%%%
% Method
%%%%%%%%%%%%%%%%%%%%%%%%%%%%%%%%%%%%%%%%%%%%%%%%%%%
\section{Method}
\label{sec:method}

This work estimates the demographics and growth of the astronomer community from the authorship
data that are available in ADS.
ADS comprises four bibliographic collections that currently contain more than 28 million records.
The `astronomy' collection is used in this work, covering astronomy, astrophysics, heliophysics and 
planetary sciences.

Using ADS, a database of authors that published articles in the period 1800--2025 was constructed
(see Appendix \ref{appendix:database} for details).
Metadata for 716\,990 publications were extracted from ADS, including author names and affiliations,
and article page numbers.
From the collected metadata, a database of authors was built that contains for each entry the author 
name and the year of the first and last time that the author appeared on a publication.
This period defines the activity period of the author.
For each year within that activity period, the database contains 
the number of co-authored papers,
the contributions to the papers, defined as the inverse of the number of co-authors summed over
all co-authored papers, 
the number of contributed pages, defined as the number of paper pages divided by the number of 
co-authors summed over all co-authored articles,
a list of affiliation identifiers, which are 6-digit long unique identifiers that are provided by ADS
and that allow associating authors to host institutes,
and a list of 2-digit ISO country codes of these host institutes.
Furthermore, the database collects information about retirement or decease years, if available.

Authors were cross-identified using a sophisticated name matching scheme that is described in
Appendix \ref{appendix:database}.
The name matching scheme imposes a maximum allowed gap in the activity period of ten
years, since it is rather unlikely that an astronomer stops publishing articles for a long period.
Authors with the same (or similar) names showing a gap of more than ten years in their publication
activity were kept as separate entries in the database.

If possible, lacking ADS affiliation identifiers were complemented by searching the affiliation texts for 
keywords that allow assigning affiliation identifiers and country codes of host institutes.
Affiliation identifiers and texts were also used to classify authors into different categories according 
to the type of their host institutions:
authors working for universities or research organisations,
authors working for space agencies,
authors working for institutes affiliated to the defence sector,
authors working for companies (including authors working for journals),
amateur astronomers,
and other authors (including authors working for museums or planetariums, libraries, astronomical
societies if they admit non-professional astronomers, foundations, think tanks or government 
organisations that do not related directly to science).

%%%%%%%%%%%%%%%%%%%%%%%%%%%%%%%%%%%%%%%%%%%%%%%%%%%
% Author database
%%%%%%%%%%%%%%%%%%%%%%%%%%%%%%%%%%%%%%%%%%%%%%%%%%%
\section{Author database}
\label{sec:database}

%%% Figure: Author database %%%%%%%%%%%%%%%%%%%%%%%%%%%%%%%%%%%
\begin{figure}[!t]
\centering
\includegraphics[width=\columnwidth]{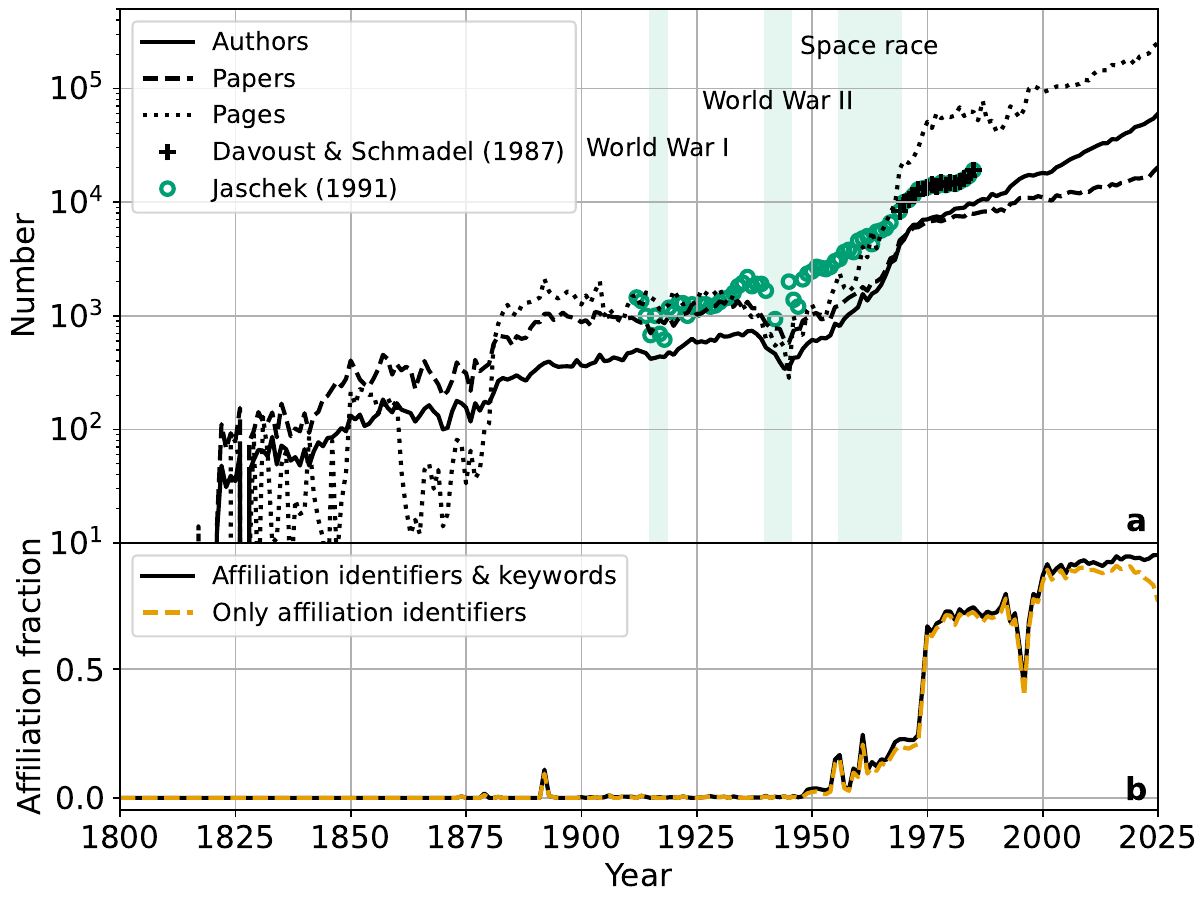}
\caption{
Number of authors, papers and paper pages ({\bf a}) and fraction of affiliated authors ({\bf b}) 
since the year 1800 in the ADS database.
Black crosses represent the number of authors given by \cite{davoust1987}, green circles are
those given by \cite{jaschek1991}.
Major geopolitical periods are indicated by light green vertical bands.}
\label{fig:database}
\end{figure}
%%%%%%%%%%%%%%%%%%%%%%%%%%%%%%%%%%%%%%%%%%%%%%%%%%%

The resulting author database contains 297\,972 authors that have published at least one paper 
within the period 1800--2025.
Panel {\bf a} of Fig.~\ref{fig:database} shows the number of authors, papers and paper pages in the 
database as function of year.
Obviously, there are little entries in the ADS database before the year 1822.
From that year on, the numbers of authors, papers and pages show a steadily increasing trend,
that is interrupted only in 1939 when the number of authors dropped significantly as a consequence 
of World War II.
World War I did not have an equivalent impact, although a modest drop in the numbers is also
visible during that period.
Note that the number of paper pages shows a more erratic trend compared to the number of
authors and papers, which may be explained by occasionally missing information on ending pages 
in ADS that affects the determination of the page numbers \citep{schulman1997}.

Following World War II the number of authors recovered and entered a phase of rapid growth.
From 1950 to 1970
the number of authors was multiplied by a factor 6,
the number of papers by a factor 5,
and the number of paper pages by a factor 19.
Interestingly, the number of papers grew slower than the number of authors, leading to an ever 
decreasing number of papers per author, a trend that was already noted by \cite{abt1981}.

Between the years 1970--1975 the growth in all numbers shows a pronounced break.
\cite{schmidtkaler1982} attributed this break to global financial difficulties, the impossibility to 
sustain a continued growth in the number of scientists, and perhaps even limits in the availability 
of talented people.
The observed continued exponential growth in the numbers since then has obviously proved 
him wrong.
For the period 1975--2025, the average growth rates were
4.3\% yr$^{-1}$ for the number of authors,
2.0\% yr$^{-1}$ for the number of papers, and
3.2\% yr$^{-1}$ for the number of paper pages.
This corresponds to doubling times of 16, 35 and 22 years, respectively.

Panel {\bf a} of Fig.~\ref{fig:database} also shows the number of authors determined by 
\cite{davoust1987} and \cite{jaschek1991}, which are clearly in excess of the numbers determined 
from ADS.
This can be explained by a larger scope of the Astronomy and Astrophysics Abstracts compared
to ADS, including also papers from neighbouring fields, solar eclipse reports and other reports
from amateur astronomers.
It is also noteworthy that the change in the growth rate around the year 1970 is less pronounced
in the data of \cite{davoust1987} and \cite{jaschek1991} compared to the ADS.
In this context it should be mentioned that author lists in ADS were truncated at the tenth author 
until about 1986 and had different amounts of truncation between 1986 and 1994 
\citep{schulman1997}.
Yet the number of papers that were concerned by these truncations is rather small, and certainly
cannot explain the magnitude of the observed differences.

Panel {\bf b} of Fig.~\ref{fig:database} shows the evolution of the fraction of authors that are 
affiliated with host institutions (and thus countries) in the database.
While the dashed line represents affiliation identifiers provided by ADS, the solid line comprises in 
addition affiliations that were derived from keywords in the affiliation texts.
Obviously, author affiliations are basically absent before 1950 and very sparse before 1975, which
may indicate a change in the management of the ADS database in that year.
The fraction of authors with affiliation identifiers evolves over the years from about $\sim60\%$ in 
1975 to $90\%$ in 2016, dropping since then to $76\%$ in 2025.
The lack of affiliations is mainly explained by a lack of affiliation texts in  ADS, with the exception
of the recent drop that is plausibly explained by a delay in ADS to assign affiliation identifiers for 
new institutes.
Deriving affiliations from affiliation text keywords compensates for this recent drop, providing
affiliations for $94\%$ of the authors in 2025.
Note that for some reason, there is a lack of affiliations in 1996, which is plausibly an artefact 
of the ADS database.

%%%%%%%%%%%%%%%%%%%%%%%%%%%%%%%%%%%%%%%%%%%%%%%%%%%
% The community in 2025
%%%%%%%%%%%%%%%%%%%%%%%%%%%%%%%%%%%%%%%%%%%%%%%%%%%
\section{The community in 2025}

%%%%%%%%%%%%%%%%%%%%%%%%%%%%%%%%%%%%%%%%%%%%%%%%%%%
\subsection{Community size}
\label{sec:communitysize}

For the year 2025 the database contains 59\,940 authors.
Table \ref{table:continent} gives their distribution per continent, together with the number of authors
per capita and per gross domestic product (GDP).
Population and GDP data were taken from the World Bank database\footnote{\url{https://data.worldbank.org}}
with GDP expressed in 2021 international dollars.
Author numbers for individual regions and countries were corrected for incomplete affiliations by 
dividing author numbers in the database by the fraction of affiliated authors in the world.

By far the largest number of authors are working in Europe, followed by Asia, North America, 
South America, Oceania and Africa.
Europe also has the largest number of authors per capita and per GDP, followed by Oceania and 
North America.
South America and Asia have a significantly smaller number of authors per capita and per GDP,
while for Africa, the number of authors per capita and per GDP is very low.

%%% Community per world region %%%%%%%%%%%%%%%%%%%%%%%%%%%%%%%%%
\begin{table}[!t]
\caption{Estimated number of authors per continent in 2025.
Authors per capita are given per million people, authors per GDP are given per billion ($10^9$) 
international 2021 US dollars.
\label{table:continent}}
\centering
\begin{tabular}{l r r r}
\hline
Region & Authors & per capita & per GDP \\
\hline
Europe & 22\,741 & 30.6 & 0.61 \\
Asia & 17\,397 &  3.6 & 0.21 \\
North America & 15\,542 & 25.4 & 0.47 \\
South America & 2\,257 &  5.2 & 0.27 \\
Oceania & 1\,168 & 25.3 & 0.59 \\
Africa & 832 &  0.5 & 0.08 \\
\hline
World & 59\,940 &  6.9 & 0.32 \\
\hline
\end{tabular}
\end{table}
%%%%%%%%%%%%%%%%%%%%%%%%%%%%%%%%%%%%%%%%%%%%%%%%%%%

%%% Figure: Community world maps %%%%%%%%%%%%%%%%%%%%%%%%%%%%%%%%
\begin{figure}[!t]
\centering
\includegraphics[width=\columnwidth]{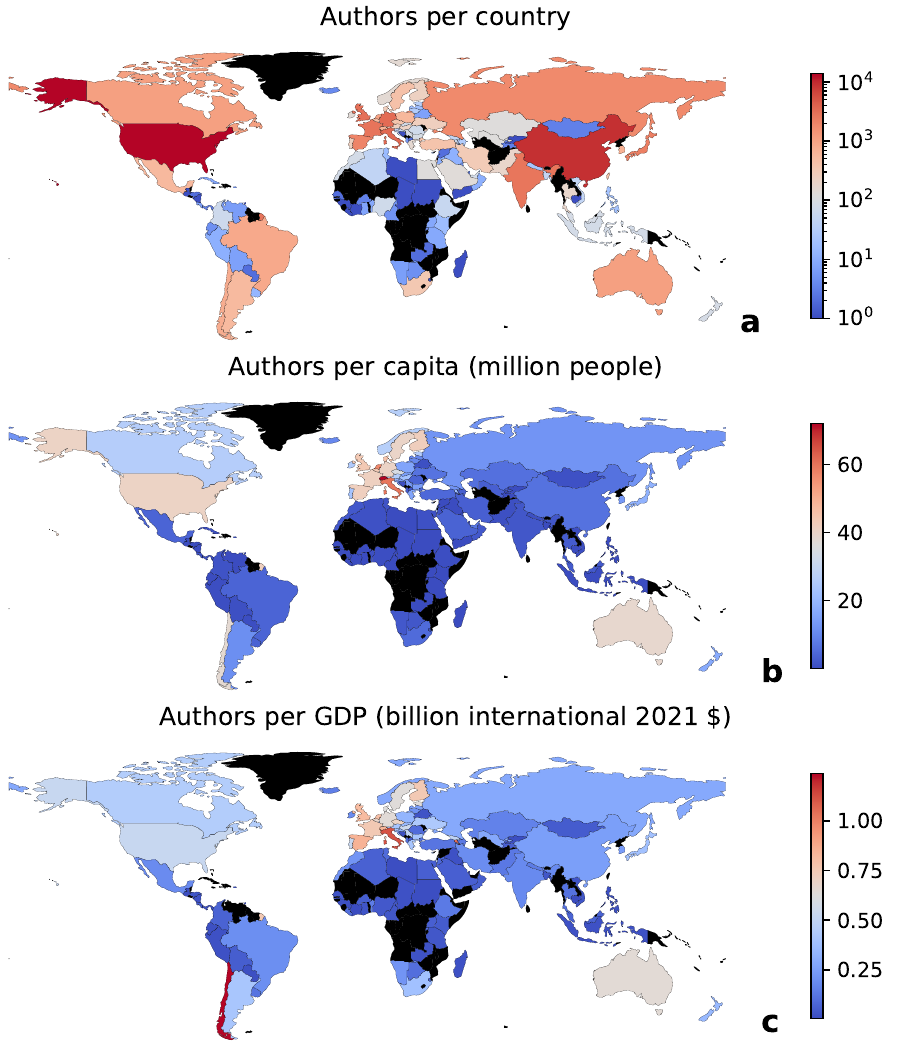}
\caption{
World maps showing the number of authors ({\bf a}, logarithmic scale), the number of authors 
per million people ({\bf b}) and the number of authors per billion ($10^9$) GDP in international 2021 
US dollars ({\bf c}) per country.
\label{fig:worldmap}
}
\end{figure}
%%%%%%%%%%%%%%%%%%%%%%%%%%%%%%%%%%%%%%%%%%%%%%%%%%%

Figure \ref{fig:worldmap} shows the same data per country in the form of world maps, corresponding
values are given in Table \ref{table:countrydata} in Appendix \ref{appendix:countrydata}.
The figure reveals that the United States is the country with the largest number of authors
(13\,865), followed by China (8\,971) and Italy (3\,526).
At the low end, there are 15 countries with only one author and 54 countries without any author
that are mostly located in Africa and Central America.

The country with the largest number of authors per million people of population is 
Switzerland (72), 
followed by Italy (60),
the Netherlands (57) 
and Denmark (46).
The United States ranks 10$^{\rm th}$ (41) and China 45$^{\rm th}$ (6), despite their large 
numbers of authors that are mainly explained by the large population sizes of these countries.

The country with the largest number of authors per billion ($10^9$) international 2021 US dollars of 
GDP is 
Chile (1.2), 
followed by Italy (1.1),
Armenia (1.0),
and Spain (0.8).
The United States ranks 19$^{\rm th}$ (0.5), behind most of the European countries and 
comparable in value to Portugal, while China is ranked 41$^{\rm th}$ (0.3), comparable
in value to Luxembourg.
The large number of authors per GDP in Chile is plausibly explained by the large density of observing 
facilities in that country that drives abundant co-authorship \citep{chen2025} and that brings
international funding that allowed for a rapid increase of astronomers in Chile \citep{reeves2023}.

%%% Figure: Category fraction maps %%%%%%%%%%%%%%%%%%%%%%%%%%%%%%%%
\begin{figure}[!t]
\centering
\includegraphics[width=\columnwidth]{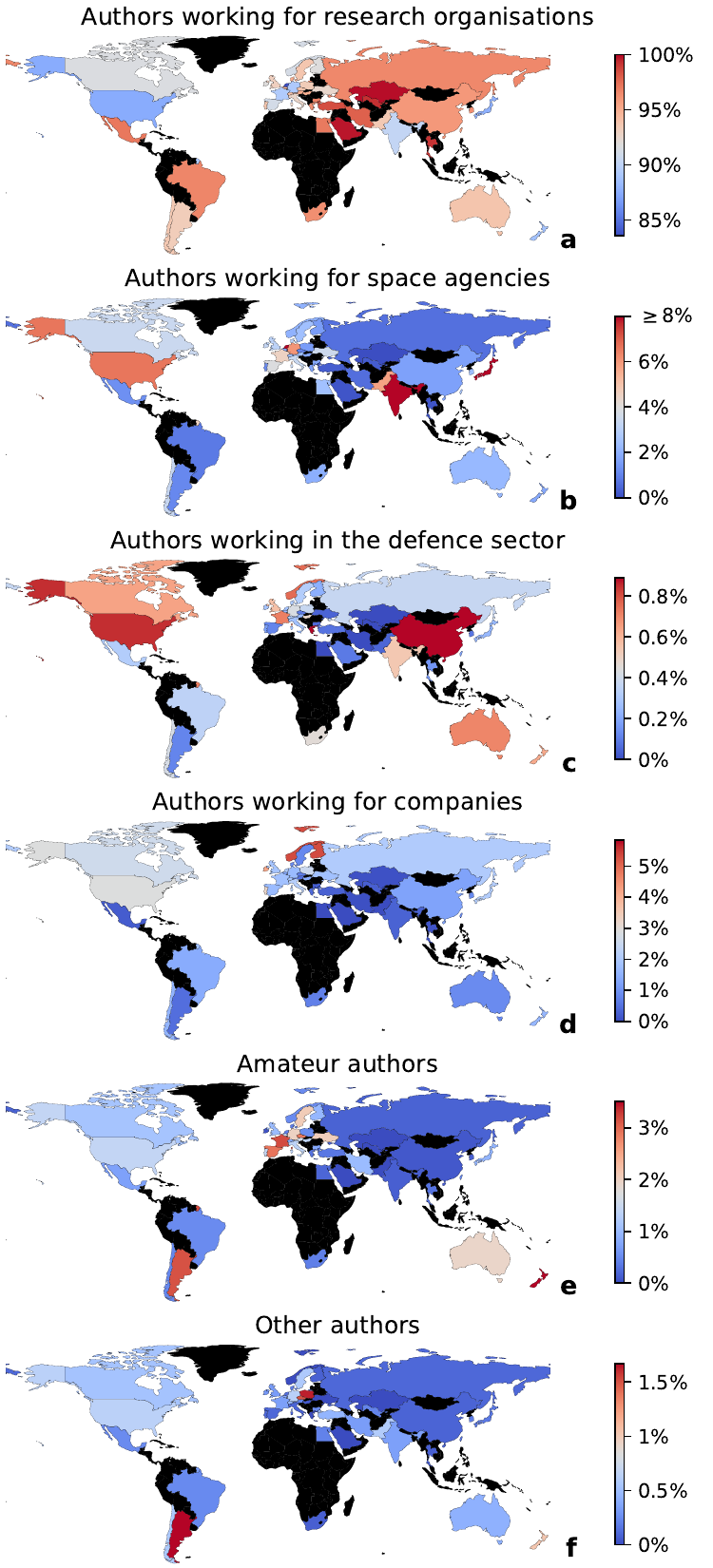}
\caption{
World maps of the percentage of authors in difference categories for the year 2025.
For space agencies, the colour scale was clipped at 8\% for a better visualisation of the variations.
Countries with less than 100 authors in 2025 are shown in black to avoid spurious results
due to small number statistics.
\label{fig:categorymaps}
}
\end{figure}
%%%%%%%%%%%%%%%%%%%%%%%%%%%%%%%%%%%%%%%%%%%%%%%%%%%

To better understand who the authors are that are publishing they were categorised according to their 
affiliations, distinguishing authors working for research organisations (including universities), space 
agencies, the defence sector, and companies, as well as amateur astronomers and other authors.
The latter category regroups authors working for museums and planetariums, libraries, astronomical 
societies that admit non-professional astronomers, foundations, think tanks or governmental 
organisations that do not relate directly to science.

Among the authors for which affiliation information was available, 
91.6\% of the authors are working for research organisations, 
4.3\% for space agencies, 
0.6\% in the defence sector, 
1.8\% for companies,
1.3\% are amateur astronomers
and 0.4\% are in the ``others'' category.
Figure \ref{fig:categorymaps} shows world maps of the author categories, illustrating that these
fractions vary considerably between countries.
Among the countries shown, those with the lowest number of authors working in research organisations
are
the Netherlands (83\%),
the United States (87\%),
Japan and New Zealand (88\%), 
Germany and France (89\%)
and India (90\%).
In the Netherlands, about 14\% of the authors work for space agencies, affiliated mainly to the European 
Space Research and Technology Centre (ESTEC) of the European Space Agency (ESA).
Other countries with significant fractions of authors working for space agencies are
Japan and India (8\%)
the United States (7\%),
and Germany (6\%).
The fractions of authors working in the defence sector are generally low, but reaches
1.3\% in Greece,
1.0\% in China,
0.9\% in the United States,
0.8\% in Norway,
and 0.7\% in France and Australia.
The fractions of authors working for companies are higher, reaching
5\% in Norway and Finland,
4\% in Ireland,
and 3\% in Portugal, the United States and Canada.
The fraction of authors who are amateur astronomers is most important in
New Zealand (7\%),
France, Argentina and Spain (3\%).
The fractions of other authors are relatively modest, but reaches
1.8\% in Argentina,
1.6\% in Poland,
1.5\% in the Czech Republic,
and 1.1\% in New Zealand.

%%% Figure: Fraction of non-publishing authors %%%%%%%%%%%%%%%%%%%%%%%%%%
\begin{figure}[!t]
\centering
\includegraphics[width=\columnwidth]{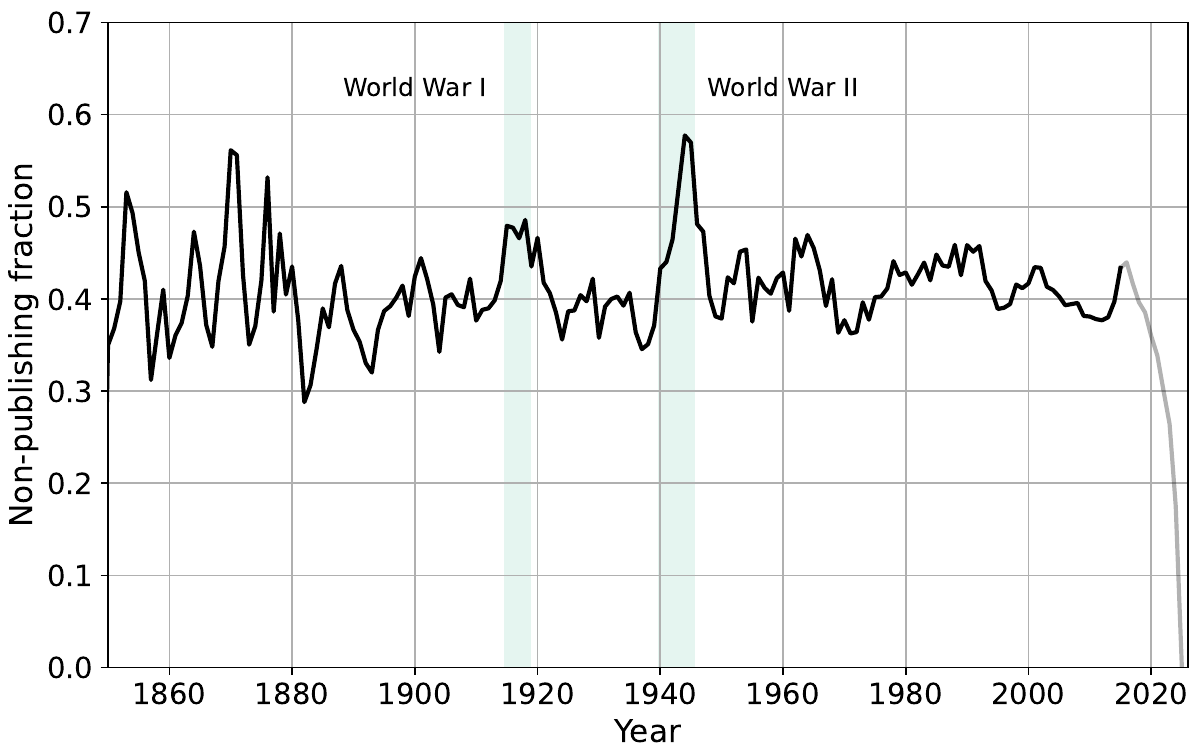}
\caption{
Fraction $f$ of non-publishing astronomers as function of year.
The drop after 2015 (shown in grey) is an artefact of the maximum activity gap of 10 years that is
applied in the post-processing of the database.
Periods of the two world wars are indicated by green vertical bands.
\label{fig:nonpublishing}
}
\end{figure}
%%%%%%%%%%%%%%%%%%%%%%%%%%%%%%%%%%%%%%%%%%%%%%%%%%%

To estimate the total size of the community the fraction $f$ of its members that are not publishing in 
a given year needs to be known.
This fraction can be estimated from the database by counting the total number of active authors 
who did not sign any paper in a given year and by dividing that number by the total number of active 
authors in that year.
An author is considered to be active if the publication year of the first and last signed paper is not 
later or earlier than the year of interest, i.e.
\begin{equation}
y^{\rm first}_i \le y \le  y^{\rm last}_i
\end{equation}
where
$y^{\rm first}_i$ and $y^{\rm last}_i$ are the years of the first and last published paper of author $i$,
respectively, and $y$ is the year of interest.
Since a maximum activity gap of ten years is applied in the post-processing, this fraction can only
by reliably estimated for years $y < 2025 -10$.

Figure \ref{fig:nonpublishing} shows how the fraction of non-publishing authors evolved over the 
years.
Overall, the fraction stays relatively constant, oscillating around a mean value of 41\%.
It shows important fluctuations in the 19th century, owing to the small number of authors in that
period, and exhibits maxima during the periods of the two world wars, reaching 58\% during
World War II.
After 2015 the fraction drops as an artefact of the post-processing of the database.
Assuming that the non-publishing fraction stays around 40\% after 2015, the total number of 
astronomers can be estimated using
\begin{equation}
N_{\rm astronomers} = \frac{N_{\rm authors}}{1-f}
\end{equation}
to about 100\,000 in the year 2025.
Note that this number is sensitive to the choice of the maximum activity gap in the post-processing
(cf.~Fig.~\ref{fig:maxgapimpact}) and hence it should be understood as a rough estimate.
A maximum activity gap of five years gives $f=0.31$ resulting in 86\,900 astronomers,
while a gap of 15 years gives $f=0.46$ resulting in 110\,000 astronomers.
The comparison of the number of astronomers to the reported scientific staff for a selected number
of institutes suggests that the derived number of astronomers is accurate to about 20\% 
(cf.~Fig.~\ref{fig:authors_institutes}).

%%%%%%%%%%%%%%%%%%%%%%%%%%%%%%%%%%%%%%%%%%%%%%%%%%%
\subsection{Age structure}
\label{sec:agestructure}

%%% Figure: Professional age distribution %%%%%%%%%%%%%%%%%%%%%%%%%%%%%
\begin{figure}[!t]
\centering
\includegraphics[width=\columnwidth]{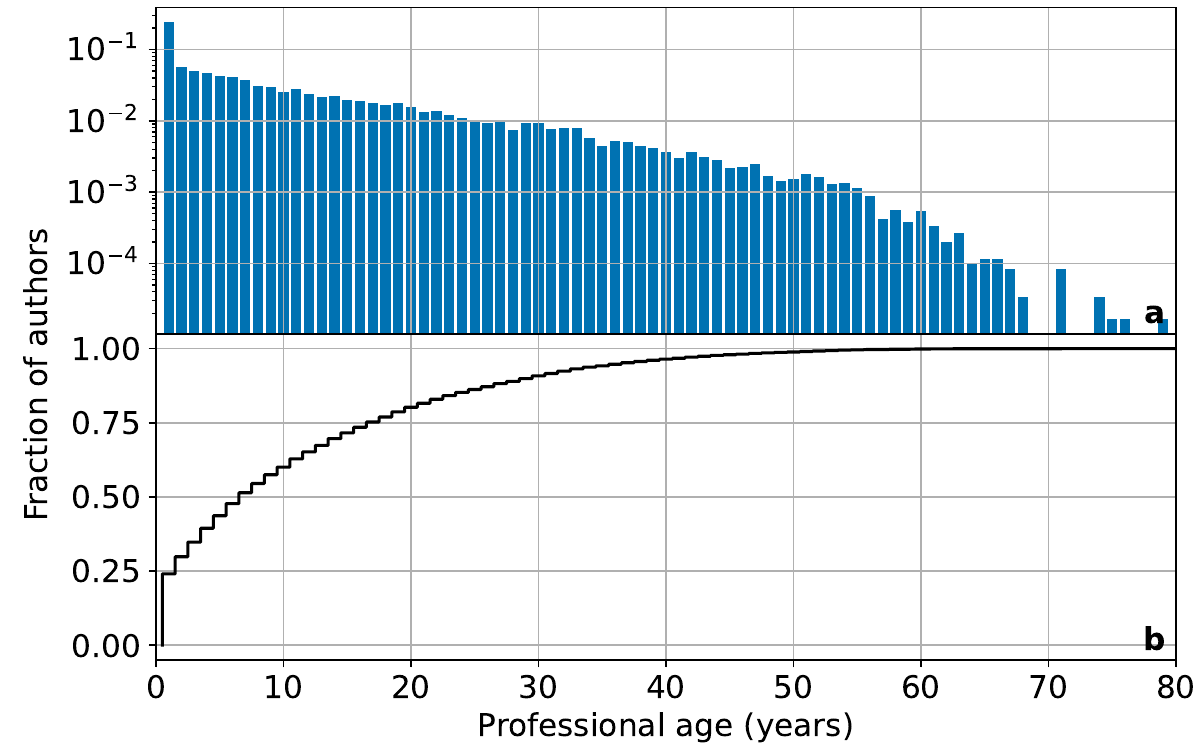}
\caption{
Professional age of the authors that were publishing in 2025.
Panel ({\bf a}) gives the fraction of authors as function of their professional age, panel ({\bf b}) gives 
the cumulative fraction of authors that have at most a given professional age.
\label{fig:agedistributions}
}
\end{figure}
%%%%%%%%%%%%%%%%%%%%%%%%%%%%%%%%%%%%%%%%%%%%%%%%%%%

The database holds also information about the professional age of the authors which is defined by 
$y^{\rm last}_i - y^{\rm first}_i +1$.
Figure \ref{fig:agedistributions} shows in panel {\bf a} a histogram of the professional ages of authors 
that were publishing in 2025.
The cumulative distribution of this histogram is shown in panel {\bf b} of Fig.~\ref{fig:agedistributions}.

The population of publishing authors is clearly dominated by early career scientists.
24\% of the authors have a professional age of one year, meaning that they were publishing for the 
first time in 2025.
Half of the authors are publishing papers since seven years or less and 75\% of the authors have
a professional age of 17 years or less.
Interestingly, there is no sharp levelling-off of the fraction of authors at a professional age where one
would expect that astronomers would retire (typically around a professional age of 35--45 years), 
which suggests that many astronomers continue publishing papers after retirement.
Note also that there are a few authors with professional ages beyond 70 years, which plausibly 
reflect name fellows that were merged into a single author by the post-processing algorithm
(cf.~Appendix \ref{appendix:database}).

\begin{figure}[!t]
\centering
\includegraphics[width=\columnwidth]{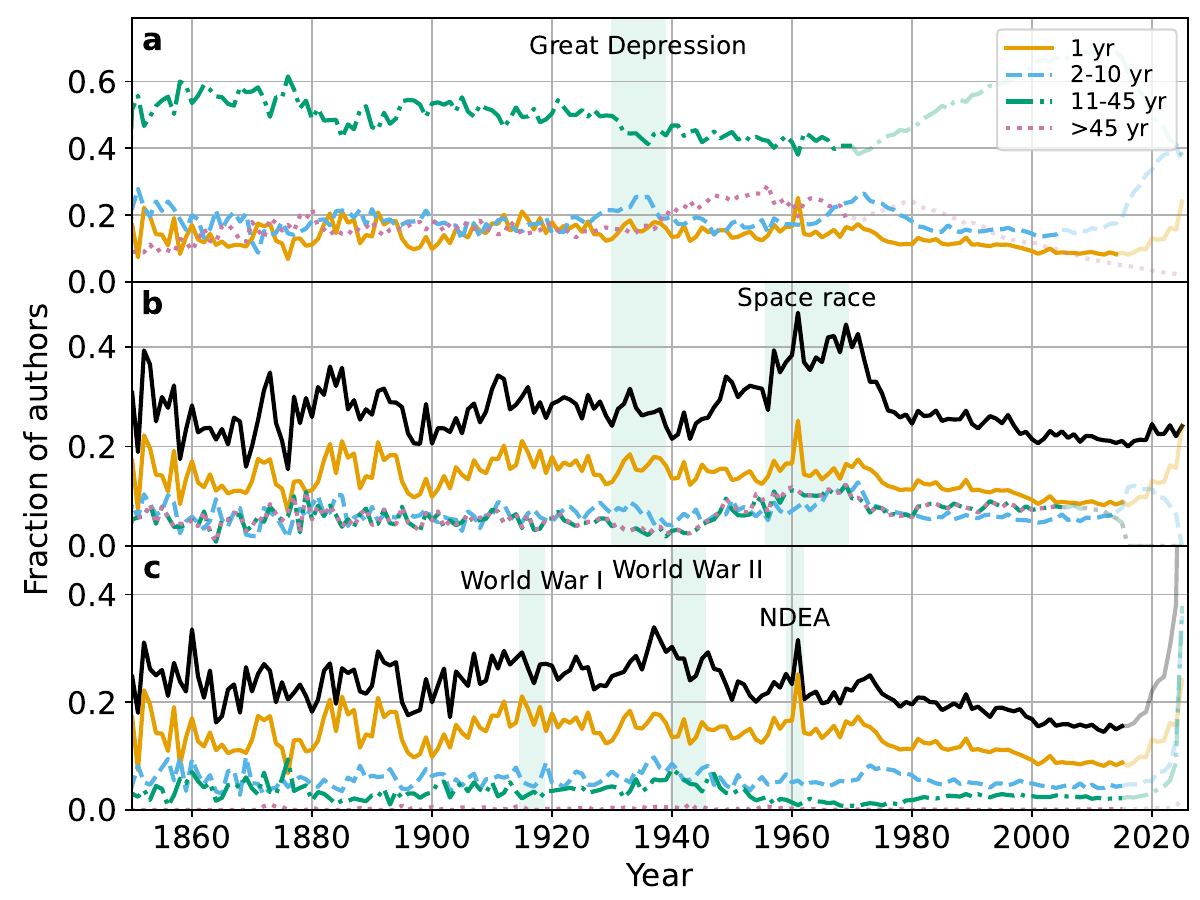}
\caption{
Fraction of authors in four different professional age classes as function of year ({\bf a}),
fraction of authors that signed their first paper in a given year ({\bf b}), and
fraction of authors that signed their last paper in a given year ({\bf c}).
The parts of the fractions that are affected by the cut-off of the database in the year 2025 and the
choice of a maximum activity gap of ten years are shown as shaded lines.
Periods of the `Great Depression', the `Space race', of increased funding provided by the 
United States National Defense Education Act (NDEA), and of the two world wars are indicated 
by green vertical bands.
\label{fig:ageclasses}
}
\end{figure}

To learn more about the age structure of the community, the authors have been classified according
to their professional age into authors that published just in a single year, and authors that published 
for 2--10 years, 11--45 years and 46--65 years.
Plausibly, these age classes correspond respectively to PhD students, post-docs, senior scientists 
and retired authors.
The fraction of authors in the different age classes are shown in Fig.~\ref{fig:ageclasses} as function
of publication year.
Note that these fractions are affected to different extents at the end of the period by the database
cut-off year of 2025, the length of the age classes intervals, and the choice of the maximum activity 
gap.
The affected periods are represented in Fig.~\ref{fig:ageclasses} by shaded lines and are ignored 
for the following discussion.

%%% Figure: Evolution of age classes over time %%%%%%%%%%%%%%%%%%%%%%%%%%
%%%%%%%%%%%%%%%%%%%%%%%%%%%%%%%%%%%%%%%%%%%%%%%%%%%

Panel {\bf a} of Fig.~\ref{fig:ageclasses} shows the evolution of the fraction of authors in the four 
different age classes.
Globally, the fractions stay relatively constant over time, with mean values of
14\% for PhD students,
18\% for post-docs,
49\% for senior scientists and
16\% for retired authors.
Yet there is a continuous decrease in the fraction of senior scientists from about 55\% in the 19th 
century to about 40\% in 1970, with a pronounced drop at the onset of the `Great Depression' in
1929.
This drop goes in hand with an increase in the fraction of post-docs, which implies that the average
professional lifetime of the authors shortened, plausibly as a cause of the rise of unemployment
during that period.
After the `Great Depression' the fraction of senior scientists increases a bit, yet the declining
trend continuous over the following years.
Interestingly, the decline in the fraction of senior scientists seems to be mainly compensated by
an increase in the fraction of retired authors.
Such an increase is expected if retired astronomers publish for increasing periods after retirement, 
which may result from the increased life expectancy at age 65 over the years \citep{un2024}.

Panel {\bf b} of Fig.~\ref{fig:ageclasses} shows the fraction of authors that published the first time in
a given year (hereafter starting authors), broken down also by age classes, while panel {\bf c} is an 
equivalent figure for the fraction of authors that published for the last time in a given year (hereafter
stopping authors).
On average (black solid curves), 28\% of the authors in a given year are starting authors while
23\% are stopping authors, resulting in an average annual net increase in the number of authors 
of about 5\%.
These fractions are subject to statistical fluctuations, in particular in the 19th century due to the 
relatively small number of authors at that time.

The fraction of starting authors exhibits a remarkable increase over the period 1945--1975, which
is also known as the `Great acceleration' that followed World War II \citep{steffen2015}.
A sharp increase in the number of starting authors is observed in 1955, which is concomitant with 
the announcement by the United States and the immediate reaction of Russia to launch satellites
into space, announcements that are considered as the starting point of the `Space race' \citep{teigens2020}.
The fraction of starting authors remained at a high level until the Moon landing in 1969 (which 
is taken as the end of the `Space race' in panel {\bf b} of Fig.~\ref{fig:ageclasses}), showing a 
significant peak in 1961 that originates exclusively from PhD students.
The number of authors that published only for a single year rose from 197 in 1960 to 391 in 1961 
and dropped back to 195 in 1962.
This peak may reflect the impact of the National Defense Education Act (NDEA) that was signed 
in 1958 in the United States in response to the launch of the Sputnik satellite and that significantly 
increased funding of education institutions \citep{flattau2006}.
Following the Moon landing, the fraction of starting authors declined, to a level that is comparable
to that observed prior to the `Space race'.

The fraction of stopping authors shows it's most important increase during the period of the
`Great Depression' (1929--1939), supporting the idea that astronomers lost their job during that
period.
Interestingly, there is no significant increase in the fraction of stopping authors during World War I,
and it's increase is limited during World War II, suggesting that most astronomers survived the wars.

%%%%%%%%%%%%%%%%%%%%%%%%%%%%%%%%%%%%%%%%%%%%%%%%%%%
\subsection{Publishing behaviour}

%%% Figure: Publications versus professional age %%%%%%%%%%%%%%%%%%%%%%%%%%
\begin{figure}[!t]
\centering
\includegraphics[width=\columnwidth]{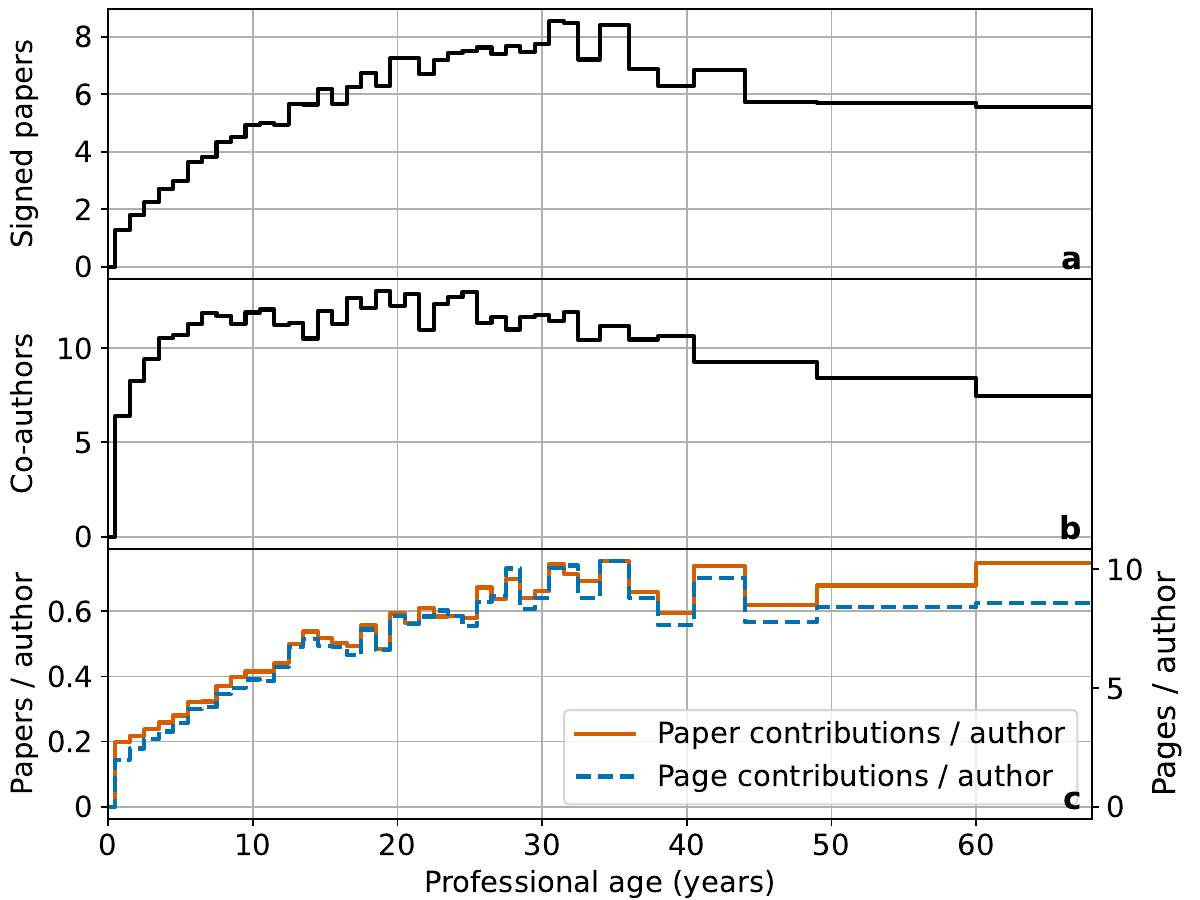}
\caption{
Average number of papers signed by an author in 2025 ({\bf a}), 
average number of co-authors on those papers ({\bf b}) and 
contributed papers and pages per author ({\bf c}) as function of the professional age of the authors.
To avoid statistical noise, the horizontal binning of the data was adjusted by requiring at least 200 
authors in each bin.
\label{fig:papersage}
}
\end{figure}
%%%%%%%%%%%%%%%%%%%%%%%%%%%%%%%%%%%%%%%%%%%%%%%%%%%

To learn more about the publishing behaviour of the community, some key numbers are shown in 
Fig.~\ref{fig:papersage} as function of the professional age of the authors for papers published in 
2025.

Panel {\bf a} shows that for scientists that signed at least one paper in 2025, the average number 
of signed papers increases steadily from a value of 1.3 for scientists that start their publishing
activity to a value of almost 8 for scientists that are active since 30 years and more.
This trend is plausibly explained by the increasing network of collaborations that astronomers
build over the years, and that increases their opportunities to sign papers.
Beyond about 35 years of activity, the average number of papers signed gradually declines, 
plausibly due to engagement in more managerial activities and retirement.

Panel {\bf b} shows that the average number of co-authors on papers evolves over the career,
raising from about 6 to around 12 over the first few years of the career, before gradually declining
after about 25 years.
The quick raise is plausibly explained by the transition from a scientific environment that is
largely limited to the PhD supervisors to a more extended network of collaborators during the
following post-doc period.
The decline is plausibly concomitant with the retirement of the authors, which may suggest a trend
that emeritus authors publish papers with less co-authors, possibly because their network of
collaborators recedes.

Panel {\bf c} shows the evolution of the number of papers and paper pages per author over the
career, recalling that `papers' and `pages' means here contributions to a paper, where the 
contribution is defined as the reciprocal number of co-authors.
The evolution of the number of papers per author thus reflects the curve of panel {\bf a} divided 
by that of panel {\bf b}.
The number of paper and page contributions per author increases steadily from the beginning
of the career until about 30 years, from which on the ratios stay about constant for the remaining
years.
While the raising part plausibly reflects the increasing network of collaborations over the years,
the constant part is a result of a declining number of papers signed and number of co-authors
that compensate each other.
Note that the number of pages per author in Fig.~\ref{fig:papersage} raises much slower at the 
beginning of the career compared to the 3--4 years to reach the plateau observed by \cite{abt1982} 
for a sample of 115 astronomers in the United States who received their PhDs in 1945--1960.

%%%%%%%%%%%%%%%%%%%%%%%%%%%%%%%%%%%%%%%%%%%%%%%%%%%
% The evolution of the community
%%%%%%%%%%%%%%%%%%%%%%%%%%%%%%%%%%%%%%%%%%%%%%%%%%%
\section{The evolution of the community}
\label{sec:evolution}

%%%%%%%%%%%%%%%%%%%%%%%%%%%%%%%%%%%%%%%%%%%%%%%%%%%
\subsection{Community size}

%%% Figure: Author evolution %%%%%%%%%%%%%%%%%%%%%%%%%%%%%%%%%%%%
\begin{figure}[!t]
\centering
\includegraphics[width=\columnwidth]{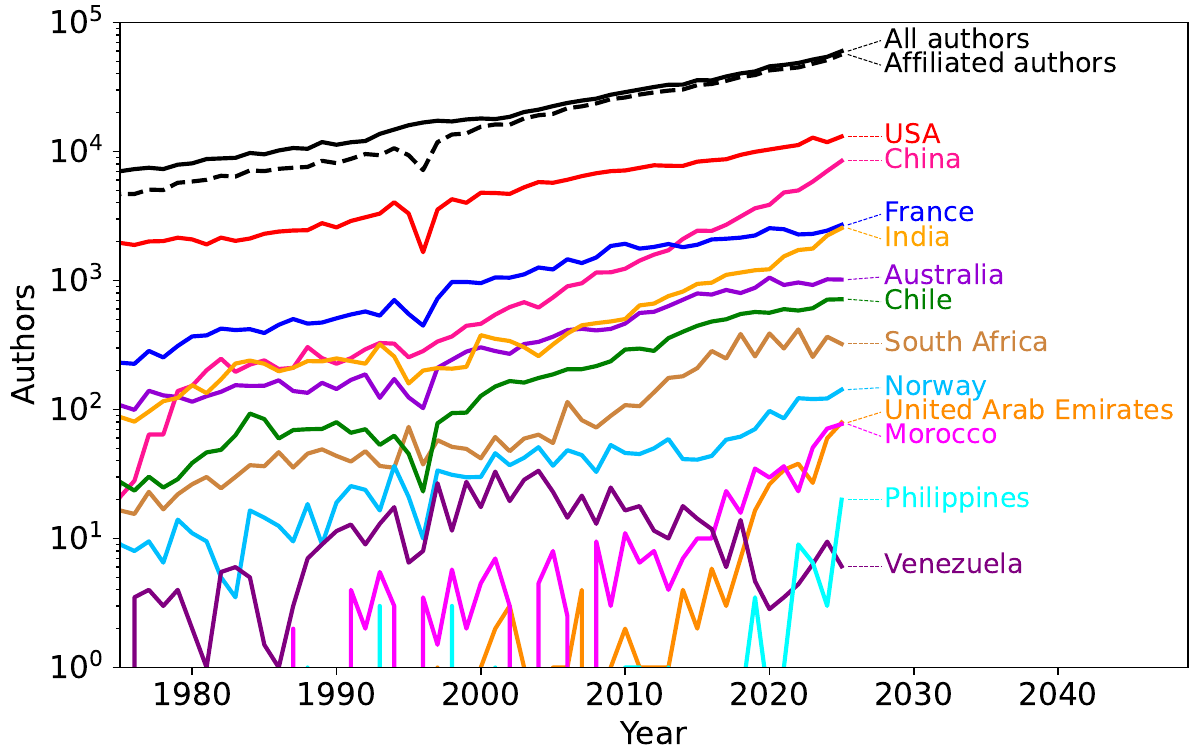}
\caption{
Evolution of the number of authors in the world and in some selected countries since 1975.
The dashed black line shows the evolution of the number of affiliated authors.
The dip in the number of authors for selected countries for the year 1996 results from missing author
affiliation information in ADS for this year.
\label{fig:authorevolution}
}
\end{figure}
%%%%%%%%%%%%%%%%%%%%%%%%%%%%%%%%%%%%%%%%%%%%%%%%%%%

The database also holds information on the evolution of the community over the years that, thanks
to the author affiliation information, can be studied on country and institute levels from the year 1975 
on.
Figure \ref{fig:authorevolution} shows the evolution of the number of authors in the world and twelve
selected countries for the period 1975--2025.
The evolution of the number of affiliated authors is also shown, which is the sum of the numbers of
all countries.
Note that this number grows a bit faster than the total number of authors, as the affiliation fraction
increases over time (cf.~Fig.~\ref{fig:database}).

The number of authors in the world increased exponentially over this period, with an average annual 
growth rate of 4.3\% yr$^{-1}$ that corresponds to a doubling-time in the number of authors of 
16 years.
This annual growth is currently speeding up, rising from 4.0\% for the period 1975--2000 to 4.7\% 
for the period 2000--2025, probably as a consequence of the strong growth of authors working in 
China and India.
Over the last five years, the average annual growth rate in both countries has reached 15\% and 
14\%, respectively, which compares to the value of 4.5\% for the United States, the country that hosts 
so far the largest community of authors.
If the United States, China and India maintain their current growth rates, the number of authors in
China and India will exceed those in the United States in 2029 and 2043, respectively.

%%% Figure: Community growth rate world maps %%%%%%%%%%%%%%%%%%%%%%%%%%
\begin{figure}[!t]
\centering
\includegraphics[width=\columnwidth]{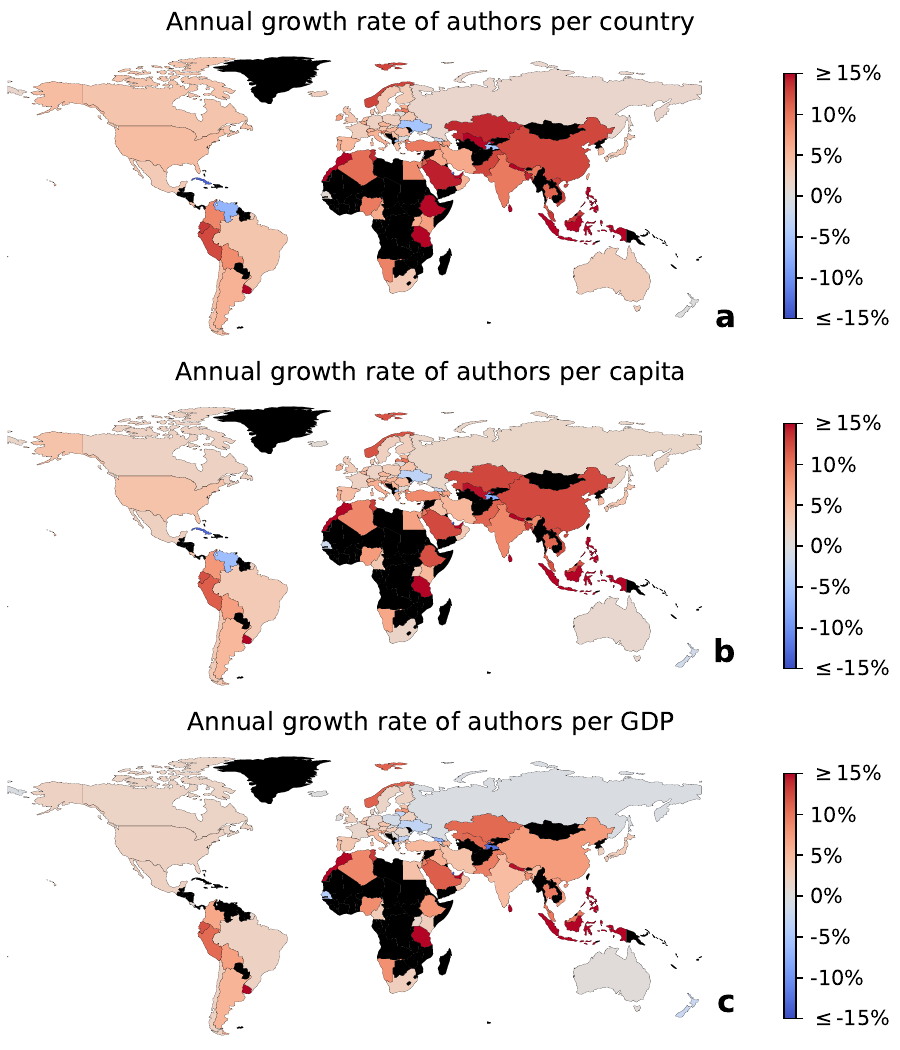}
\caption{
World maps showing the annual growth rate in the number of authors ({\bf a}), the number of authors
per capita ({\bf b}) and the number of authors per GDP ({\bf c}) per country.
The growth rates were determined over the last ten years (2015--2025).
Colour scales were clipped at $\pm$15\% for a better visualisation of the variations.
Countries with less than 30 authors over the last ten years are shown in black.
\label{fig:growthmaps}
}
\end{figure}
%%%%%%%%%%%%%%%%%%%%%%%%%%%%%%%%%%%%%%%%%%%%%%%%%%%

Figure \ref{fig:growthmaps} shows world maps of the growth rates of the number of authors, authors
per capita and authors per GDP that were determined for the period 2015--2025.
Only countries which host at least 30 authors over that period were displayed to avoid fluctuations 
from small number statistics.
Numerical values of the growth rates are provided in Table \ref{table:countrydata} of 
Appendix \ref{appendix:countrydata}.

While growth rates vary strongly among countries, the number of authors, authors per capita and 
authors per GDP has grown in almost all countries over the last ten years.
The country with the largest growth rate in the number of authors are 
the Philippines (42\%),
followed by
the United Arab Emirates (35\%), 
Luxembourg (24\%),
Indonesia (22\%), 
and Uruguay (21\%).
All these countries have in 2025 a few tens authors at most, hence they may be considered as 
emerging countries in the field of astronomy.

There are only eight countries in which the number of authors decreased, including
Cuba,
Georgia,
Lebanon,
North Macedonia,
Qatar,
Tajikistan,
Ukraine, and
Venezuela.
In most of these countries the decrease is plausibly explained by wars or severe economic and 
political crises.

%%%%%%%%%%%%%%%%%%%%%%%%%%%%%%%%%%%%%%%%%%%%%%%%%%%
\subsection{Publishing behaviour}

%%% Figure: Evolution of the community output %%%%%%%%%%%%%%%%%%%%%%%%%%
\begin{figure}[!t]
\centering
\includegraphics[width=\columnwidth]{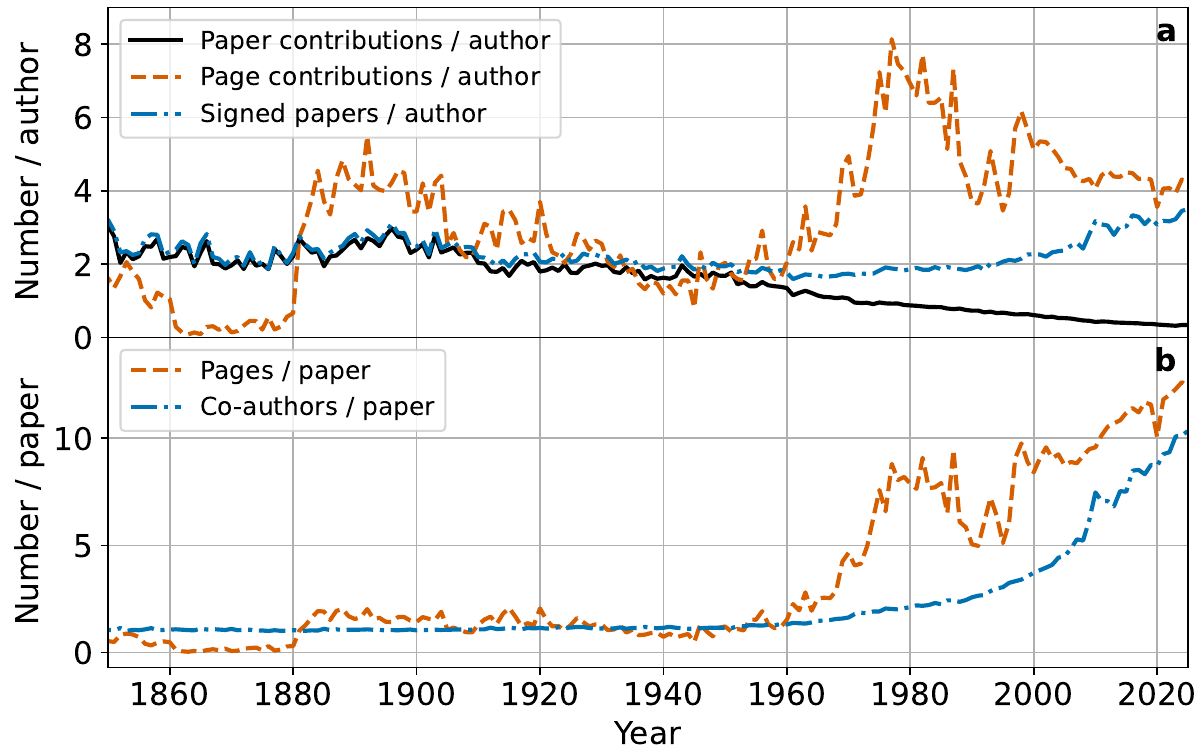}
\caption{
Evolution of the publishing behaviour of the community for the period 1850--2025.
Panel {\bf a} shows the evolution of the annual number of paper contributions (black solid line) and 
page contributions (orange dashed line) per author together with the average number of papers 
signed per author and year (blue dashed-dotted line).
Panel {\bf b} shows the evolution of the number of pages and co-authors per paper.
\label{fig:output}
}
\end{figure}
%%%%%%%%%%%%%%%%%%%%%%%%%%%%%%%%%%%%%%%%%%%%%%%%%%%

To understand how the publishing behaviour of the community evolved over the years, the evolution 
of some key numbers is shown Fig.~\ref{fig:output} over the period 1850--2025, with numbers per 
author plotted in panel {\bf a} and numbers per panel plotted in panel {\bf b}.

Figure \ref{fig:output} shows that the annual number of paper contributions per author (panel {\bf a}, 
black solid line), defined as the total number of papers published in a given year divided by the 
total number of authors, is on steady decline.
In other words, individual astronomers are publishing less and less papers.
Per author, the average number of papers has declined from over 2 papers in the 19th century to
0.3 papers in 2025.
Since the number of astronomers is estimated to be about 40\% larger than the number of authors 
in a given year (cf.~section \ref{sec:communitysize}), the average number of papers per astronomer 
is correspondingly lower.

This declining trend was already observed by \cite{abt1981} in the 1980s for American astronomical
publications, who also noticed that the paper length increased between 1910 and 1980.
Figure \ref{fig:output} confirms this increase (panel {\bf b}, orange dashed line), but suggests at the 
same time a drop in the number of pages between 1987 and 1996.
Whether this drop is an artefact of the ADS database or reflects a real change of the publication
behaviour over this period remains to be understood.
In any case, the drop is followed by a continued increase that reached in 2025 an average value of 
about 12 pages per paper.
As a result, the number of page contributions per author (panel {\bf a}, orange dashed line),
defined as the total number of pages published in a given year divided by the total number of 
authors, shows a more erratic behaviour.
While the average number of pages published per author has increased over the period 
1960--1977, reaching a maximum of about 8 pages per author in 1977, it declined since then to
a value of 4 pages per author in 2025.

Figure \ref{fig:output} also shows that the average number of co-authors per paper (panel {\bf b}, 
blue dashed-dotted line) has steadily increased over the years.
While up to 1960 this number was close to one, which means that most papers were single-author
papers, the number of co-authors per paper has grown to an average value of about 10 in 2025.
This increasing trend was also already noted by \cite{abt1981}, but has accelerated significantly
since then.
As a result, the average number of papers signed by an author in a given year is increasing 
(panel {\bf a}, dashed-dotted line), and went from about 1.6 in 1965 to about 3.4 in 2025.

%%%%%%%%%%%%%%%%%%%%%%%%%%%%%%%%%%%%%%%%%%%%%%%%%%%
% Evolution of other research disciplines
%%%%%%%%%%%%%%%%%%%%%%%%%%%%%%%%%%%%%%%%%%%%%%%%%%%
\section{Evolution of other research disciplines}
\label{sec:disciplines}

%%% Figure: Comparison of author databases %%%%%%%%%%%%%%%%%%%%%%%%%%%
\begin{figure}[!t]
\centering
\includegraphics[width=\columnwidth]{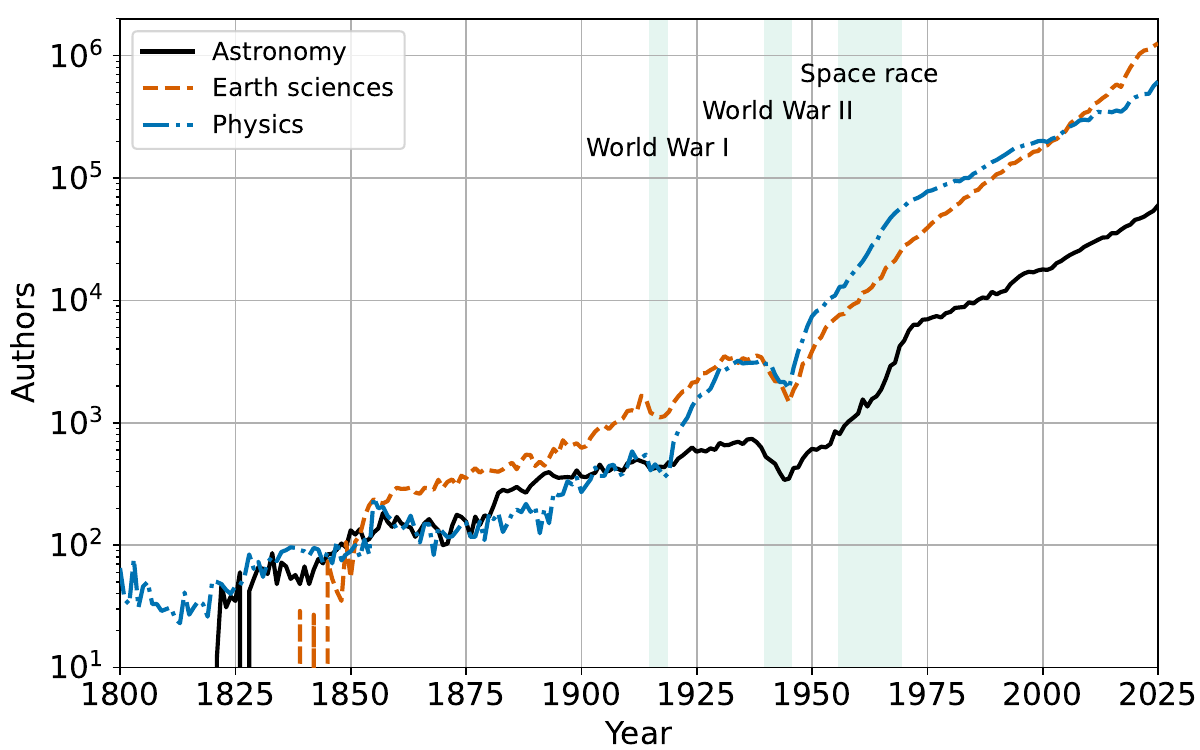}
\caption{
Number of authors for the astronomy (black solid line), earth sciences (orange dashed line) and 
physics collections (blue dashed-dotted line).
Major geopolitical periods are indicated by green vertical bands.
\label{fig:disciplines}
}
\end{figure}
%%%%%%%%%%%%%%%%%%%%%%%%%%%%%%%%%%%%%%%%%%%%%%%%%%%

To investigate whether the strong growth in the number of authors is unique to astronomy, author
databases were also constructed using the procedure described in section \ref{sec:method} and 
Appendix \ref{appendix:database} for the earth sciences and physics collections of ADS.
Figure \ref{fig:disciplines} compares the evolution of the number of authors over
the years for these two collections to those of the astronomy collection (cf.~Fig.~\ref{fig:database}).

During the 19th century, the number of authors in the three disciplines was roughly of the same 
order, although they grew at different rates.
While the number of authors in astronomy grew over the period 1850--1900 at average annual 
rate of 2.6\%, average growth rates in earth sciences and physics were 3.0\% and 1.6\%,
respectively.
Interestingly, the number of authors in physics even declined in the period 1855--1870, which
explains the relatively small growth rate over the period 1850--1900.

This situation changed dramatically in the 20th century, with physics becoming the most rapidly
growing discipline.
Over the period 1900--1950 physics grew at an average annual rate of 6.1\% yr$^{-1}$ while
earth sciences continued with 2.9\% yr$^{-1}$ the trend of the 19th century.
On the other hand, with an average annual growth rate of only 0.6\% yr$^{-1}$, the astronomy 
community grew only slowly over the period.
In 1950, as result of these different growth rates, the numbers of authors in earth sciences and 
physics have largely surpassed that in astronomy, reaching about an order of magnitude more 
authors.

Interestingly, while the number of authors in earth sciences and physics started a sustained 
growth after World War I that was only interrupted by World War II, the strong growth of the
astronomy community only started by 1955, concomitant with the beginning of the `Space race'
(cf.~section \ref{sec:agestructure}).
The growth in the number of astronomers slowed down markedly in 1970, right at the moment 
when the United States landed men on the Moon, while no comparable slow down is observed
in earth sciences and physics at that moment.
This further supports the idea that the `Space race' was a major driver of the growth of the 
astronomy community.

Over the period of the `Great acceleration' (1945--1975), astronomy, earth sciences and physics
reached average annual growth rates of 10.2\% yr$^{-1}$, 10.0\% yr$^{-1}$ and 11.3\% yr$^{-1}$,
respectively, which explains why despite the rapid growth during the `Space race', the number 
of authors in astronomy did not catch up with those in the other two disciplines.
While the growth of physics slowed down around the year 1965, the growth in earth sciences
is continuous, with made the latter the leading community of the three disciplines from the year 
2009 on.

Following the period of the `Great acceleration', average growth rates over the period 1975--2025 
are relatively stable, and established at values of
4.3\% yr$^{-1}$ for astronomy, 
6.7\% yr$^{-1}$ for earth sciences 
and 4.0\% yr$^{-1}$ for physics.
This demonstrates that the observed growth in the number of authors is not unique to astronomy, but
rather at the low side compared to other disciplines of natural sciences.

%%% Figure: Discipline output %%%%%%%%%%%%%%%%%%%%%%%%%%%%%%%%%%%
\begin{figure}[!t]
\centering
\includegraphics[width=\columnwidth]{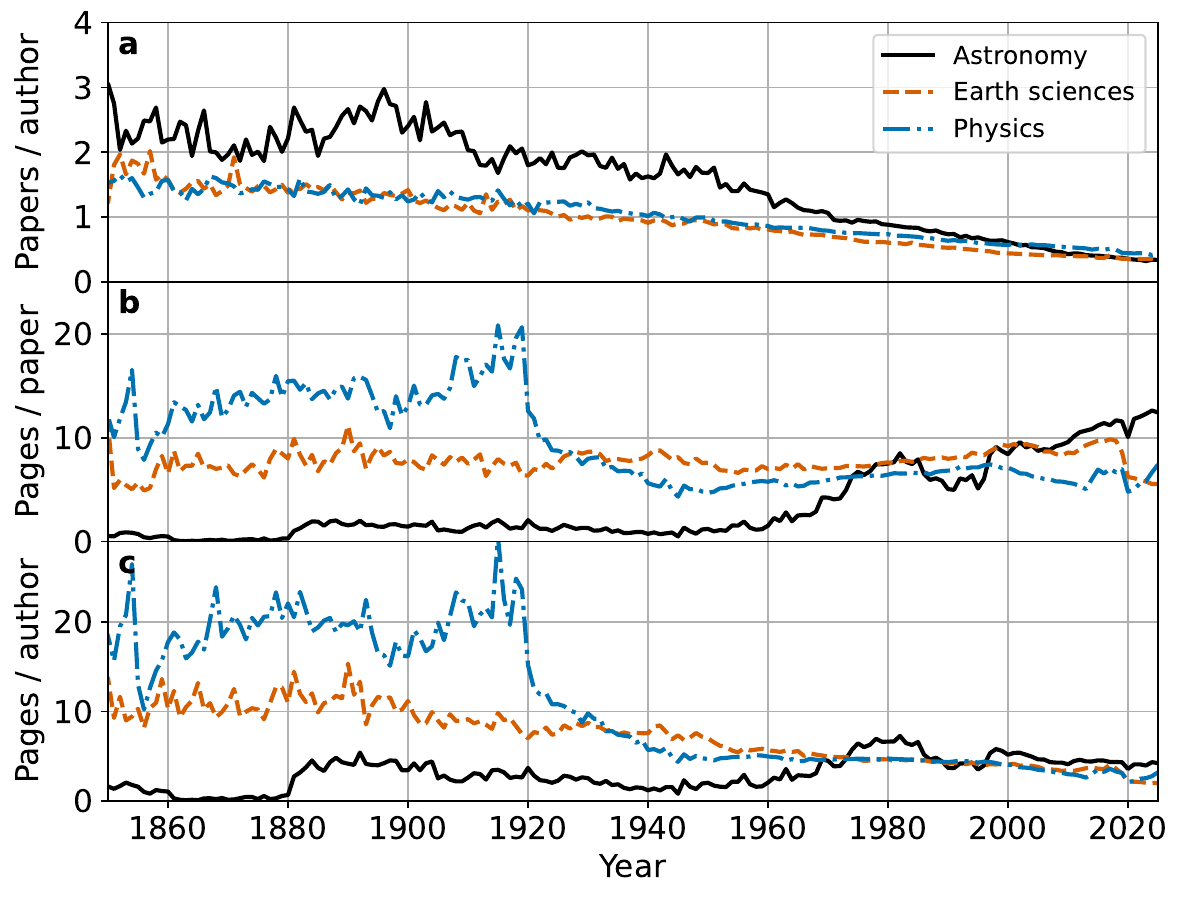}
\caption{
Evolution of the annual number of contributed papers per author ({\bf a}), pager per paper ({\bf b})
and contributed pages per author ({\bf c}) since 1850 for astronomy (black solid lines), earth sciences 
(orange dashed lines) and physics (blue dashed-dotted lines).
\label{fig:disciplineoutput}
}
\end{figure}
%%%%%%%%%%%%%%%%%%%%%%%%%%%%%%%%%%%%%%%%%%%%%%%%%%%

To understand whether the publication behaviour in earth sciences and physics differs from that in
astronomy, Fig.~\ref{fig:disciplineoutput} compares some key numbers of these disciplines to those
of astronomy.

Panel {\bf a} shows that also in earth sciences and physics, the number of papers per author is 
decreasing with time, starting at even lower values compared to astronomy in the 19th century.
The number of papers per author in earth sciences is reaching a value of about 0.4 in 2025, which
is slightly in excess to the value of 0.3 observed for astronomy.
The evolution of the number of papers per author follow a very similar trend for earth sciences
and physics, with a decrease that is significantly slower than that observed for astronomy.

Panel {\bf b} shows that the number of pages per paper evolved quite differently for the different
disciplines.
While for earth sciences this number was rather stable around a value of 8 over most of the time (and 
decreased only significantly over the last five years), for physics this number was increasing until
1919 to over 20 pages per paper, followed by a rapid drop to a value of about 6 that was relatively
stable over the past 80 years.
This is in contrast to the increasing number of pages per paper in astronomy, that in 2025 has 
reached about twice the value observed in earth sciences and physics.
Astronomy is hence the only of the three disciplines where the number of pages per paper increased
significantly over the past 50 years.

As shown in panel {\bf c}, this growth compensates the decrease in the number of papers per author,
making astronomy the discipline that produces today the largest number of pages per author.
Still, in all disciplines, also the number of pages per author show a decreasing trend, reaching
in 2025 values of about 4, 3, and 2 for astronomy, physics and earth sciences, respectively.

%%%%%%%%%%%%%%%%%%%%%%%%%%%%%%%%%%%%%%%%%%%%%%%%%%%
% Discussion
%%%%%%%%%%%%%%%%%%%%%%%%%%%%%%%%%%%%%%%%%%%%%%%%%%%
\section{Discussion}

The analysis of bibliographic information available in ADS has unveiled the sustained and exponential
growth of the astronomy community in the world.
While this growth has already been recognised by \cite{schmidtkaler1982}, \cite{davoust1987} and 
\cite{jaschek1991} in bibliographic data up to the year 1985, this is the first time that a continued
and unattenuated growth is document up to the year 2025.
\cite{davoust1987} estimated that the number of astronomers in the world in 1985 was comprised
between 10\,000 and 25\,000, this study reveals that it has reached about 100\,000 $\pm$ 20\,000 
individuals in 2025.

A large fraction of the community is working in Europe and North America, with Asia currently
catching up, owing to the strong growth of the number of astronomers in China and India.
Over the last five years, annual growth rates have reached 15\% (China) and 14\% (India), which 
implies that their astronomer communities currently double every five years.
At that pace, China will host the largest astronomer community in the world by 2029, while India
will follow second in the year 2043.
A continuous growth is probably relatively easy to sustain in both countries since the number of
astronomers per capita of population or per GDP are still well below the levels that are found
in Europe or in Northern America.
Therefore, the barycentre of astronomy will probably shift from Europe and Northern America
to Asia, driven by the large populations in this region and the apparent political will to develop 
astronomy.

%%% Figure: Space race %%%%%%%%%%%%%%%%%%%%%%%%%%%%%%%%%%%%%
\begin{figure}[!t]
\centering
\includegraphics[width=\columnwidth]{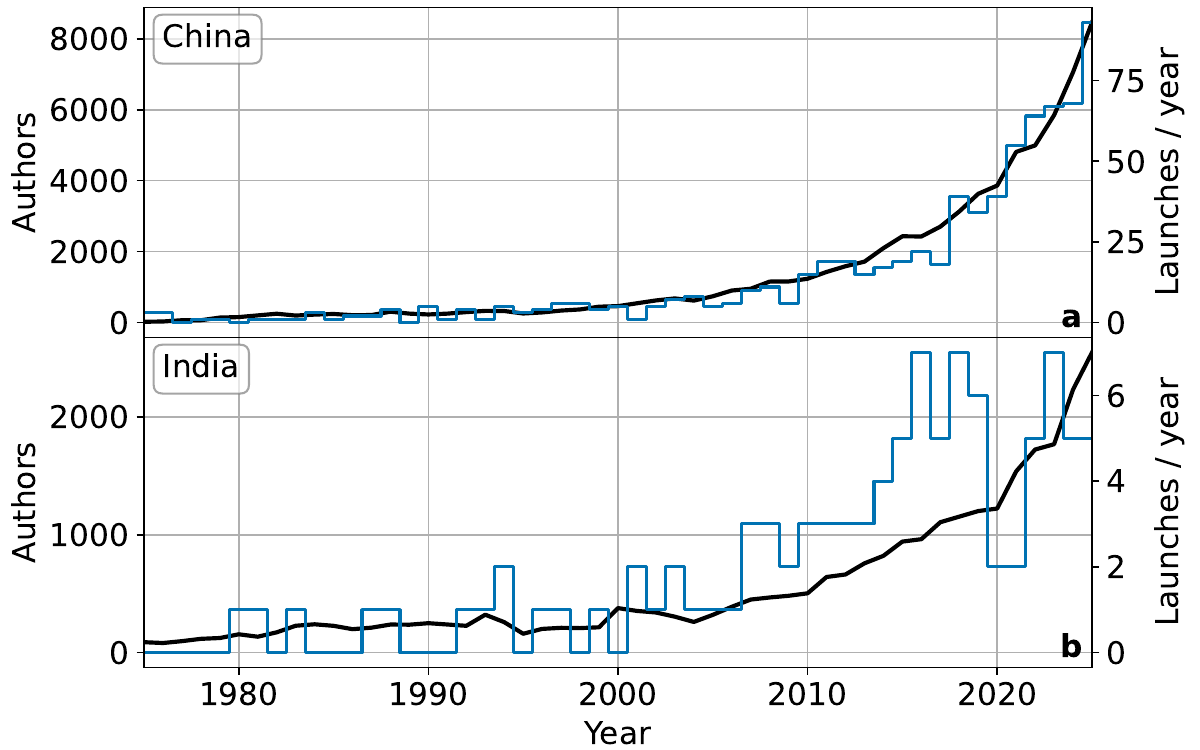}
\caption{
Evolution of the number of authors and annual rocket launches in China ({\bf a}) and India ({\bf b}) 
since 1975.
Rocket launch data were taken from \url{https://spacestatsonline.com/launches/country/}.
Plots are in linear scale.
\label{fig:spacerace}
}
\end{figure}
%%%%%%%%%%%%%%%%%%%%%%%%%%%%%%%%%%%%%%%%%%%%%%%%%%%

This study presents evidence that one of the drivers for this political will is the desire to compete in
the `Space race'.
The catching up of the United States in the first `Space race' is a plausible explanation of the rapid 
expansion of the astronomer community in the period 1955--1970.
The `New space race', that is currently unfolding, is more multipolar \citep{reed2026}, with China 
and India trying to catch up, but including also countries like the United Arab Emirates or Saudi 
Arabia.
It is plausible that the `New space race' is driving the growth of the astronomer community in these
countries.
To support this hypothesis, the evolution of the number of authors is compared in 
Fig.~\ref{fig:spacerace} to the evolution of the number of rocket launches per year in China and 
India.
At least for China, there exists a close co-evolution of the number of authors and the number of 
rocket launches, with an average number of about 100 authors per rocket launch over the period
2010--2025.
The data for India do not exhibit such a close co-evolution, but as the number of annual rocket 
launches is still low in this country, the latter is subject to statistical variations.
Still, the rapid growth of the number of authors in India started at about 2005 when the number of 
rocket launches also increased, and as shown in section \ref{sec:communitysize}, 8\% of these 
authors work directly for the Indian space agency in 2025.
It is therefore of little doubt that the growth of the community in India is also driven by the will
to compete in the `New space race'.
Finally, while the United Arab Emirates has so far not launched any rocket into space, the current
annual author growth rate of 35\% goes in hand with an ambitious space program, which made
it the first Arab country to orbit Mars, to launch a lunar rover and to run an astronaut programme 
\citep{alfarsi2026}.

Another interesting insight from this study is that the publication activity of the astronomers is
continuously decreasing over time, be it in the number of papers or in the number of pages 
published.
The exponential literature growth in astronomy is hence clearly not the result of an increased 
publication activity but the result of the growth of the community.
This observation was already made by \cite{schmidtkaler1982} and \cite{davoust1987} in the
1980s, yet it is interesting to observe that since then the trend continued.

It is also interesting to note that the number of astronomers grows faster than the number of
research infrastructures that are available for astronomy.
\cite{knodlseder2024} determined growth rates of 1\% yr$^{-1}$ and 3.2\% yr$^{-1}$ for ground-based 
and space facilities over the last 30 years, respectively, which are smaller than the observed growth 
rate of 4.3\% yr$^{-1}$ in the number of authors.
This implies that less and less research infrastructures are available per astronomer, and suggests
that a major driving force in their increase is the strong growth of the astronomer community.

%%%%%%%%%%%%%%%%%%%%%%%%%%%%%%%%%%%%%%%%%%%%%%%%%%%
% Conclusions
%%%%%%%%%%%%%%%%%%%%%%%%%%%%%%%%%%%%%%%%%%%%%%%%%%%
\section{Conclusions}

This work has provided up-to-date information on the demographics and growth of the astronomer
community using bibliographic information available in ADS.
The most striking result of this study is the continuous and strong growth of the community that is
plausibly driven by geopolitical considerations, and specifically by the will of nations to develop and
sustain space capacities.
The strong growth is not unique to astronomy, but also observed for earth sciences and physics, which
exhibit even larger growth rates.
While speculating about the origin of the growth rates in these fields is beyond the scope of this paper,
it seems plausible that geopolitical interests also play an important role there, specifically for the field
of earth sciences that are relevant to secure access to natural ressources. 

The study has also revealed some facts that are inconsistent with popular narratifs.
The data do not show any evidence for a decline of the community, except for a handful of countries 
that currently struggle with war or economic and political difficulties.
In almost all countries, the astronomer community grows faster than the population and the GDP.
The data do also show no evidence for an increased publication activity of astronomers, the contrary
is actually true.
Astronomers are publishing less and less papers, and also the number of pages published every
year by an astronomer are on decline.

It would be interesting to understand why the narratifs differ so much from the evidence.
I can only speculate on the reasons behind this.
Maybe the available funding is rising slower than the number of astronomers, which would imply
that less and less money would be available for each individual to do research.
The slower growth rate of the research infrastructures actually points into that direction.
Astronomers could therefore feel increasingly constrained, mimicking a situation of scarcity in a
research discipline that is characterised by abundance.

%%%%%%%%%%%%%%%%%%%%%%%%%%%%%%%%%%%%%%%%%%%%%%%%%%%
% Acknowledgments
%%%%%%%%%%%%%%%%%%%%%%%%%%%%%%%%%%%%%%%%%%%%%%%%%%%
\section*{acknowledgements}
No AI was used in the design and execution of this work.
This work has made use of the 2D plotting Python library matplotlib \citep{hunter2007},
the geopandas Python library \citep{jordahl2020},
and vector map data from Natural Earth (\url{https://www.naturalearthdata.com}).
This research has made use of the Astrophysics Data System (ADS), funded by NASA under 
Cooperative Agreement 80NSSC21M0056.
The author wants to thank Jeffrey Pomerantz for his help with the ADS system during this study.

%%%%%%%%%%%%%%%%%%%%%%%%%%%%%%%%%%%%%%%%%%%%%%%%%%%
% References
%%%%%%%%%%%%%%%%%%%%%%%%%%%%%%%%%%%%%%%%%%%%%%%%%%%
\bibliographystyle{aa}
\typeout{}
\bibliography{references}

%%%%%%%%%%%%%%%%%%%%%%%%%%%%%%%%%%%%%%%%%%%%%%%%%%
% Appendices
%%%%%%%%%%%%%%%%%%%%%%%%%%%%%%%%%%%%%%%%%%%%%%%%%%
\begin{appendix}

%%%%%%%%%%%%%%%%%%%%%%%%%%%%%%%%%%%%%%%%%%%%%%%%%%%
% Creation of the database
%%%%%%%%%%%%%%%%%%%%%%%%%%%%%%%%%%%%%%%%%%%%%%%%%%%
\section{Creation of the database}
\label{appendix:database}

For this work, a database of authors that were signing papers in the period 1800--2025 was created
using the ADS digital library.
In a first step, the metadata of all refereed articles were retrieved on an annual basis for the years 
1800--2025.
The inclusion of non-refereed articles in the database did not alter the results of this study, but it led
to unreasonable paper page numbers, and hence I excluded these articles from the database.

ADS comprises four bibliographic collections covering publications in astronomy and astrophysics,
physics, earth science, and general science.
Only articles that belong exclusively to the astronomy and astrophysics collection were selected, 
since articles that are also attributed to other collections tend to reach beyond the astronomy 
community, which would lead to an overestimation of its size.
This resulted in metadata for 716\,990 publications.

For each publication the metadata retrieved included the title, the page numbers, and the names 
and affiliations of all authors.
Note that author lists in ADS were truncated at the tenth author until about 1986 and had different 
amounts of truncation between 1986 and 1994 \citep{schulman1997}, yet the fraction of the papers
concerned by this truncation is low.
As authors often sign multiple papers in a given year their name will likely emerge even if it got
truncated on some papers, hence the impact of the truncation on the results presented in this
paper should be minor.

ADS complements affiliations with 6-digit long unique identifiers that allow associating authors to
specific host institutes.
By the time of doing the analysis for this paper, 7\,776 different affiliations identifiers were supported 
by ADS.\footnote{\url{https://github.com/adsabs/CanonicalAffiliations/blob/master/country_parent_child.tsv}}
Note that authors in ADS are not cross-matched among publications, and affiliation identifiers are not 
provided for every author.
The affiliation identifiers provided by ADS were complemented by a search for keywords in the
affiliation texts that allows to identify additional author host institutions.
As panel {\bf b} of Fig.~\ref{fig:database} shows, this increased the fraction of affiliated authors, in 
particular for the most recent years where ADS affiliations are less complete.

Affiliation identifiers and keywords were also classified by the nature of the host institution,
distinguishing between
universities or research organisations,
space agencies,
institutions affiliated to the defence sector, and
companies.
Furthermore, amateur astronomers were identified in this process, sometimes through their private 
home address.
Finally, an `other' category includes
museums or planetariums, 
libraries, 
astronomical societies if they admit non-professional astronomers, 
foundations,
think tanks, 
and government organisations that are not directly dealing with science.
On the basis of these distinctions, one or several categories were assigned to each affiliated
author.

From the collected metadata, a database of authors was built that contains for each entry 
the author name,
the first and last year that the author name appeared on a publication, 
the years of retirement and decease of the author,
and the author category.
Author names were normalised by splitting them into surname and given names and reassembling 
the names by making sure that the surname always precedes the given names.
Furthermore, all unicode escape sequences were replaced by ASCII characters using a transliteration 
scheme and the names were converted to upper case.
Retirement and decease years were determined from specific affiliation identifiers and 
keywords, yet the number of authors for which this information is available is small.
The years of the first and last publication define the activity period of the author, and for
each year in this period, the database contains
the number of co-authored publications,
the contributions to the articles,
the affiliations identifiers,
and the 2-digit ISO country codes of the host institutions that were derived from these identifiers, 
taking into account the possibility of multiple affiliations.
Author contributions were defined as the reciprocal of the number of co-authors.
For example, a 10 pages article that was signed by 5 authors will be counted as an article
contribution of 0.2 and a page contribution of 2.

In building the database, authors were only merged if they had identical normalised names,
their given names were not abbreviated, and if their annual publication activities show no
temporal gap.
This largely avoids the merging of name fellows while keeping the size of the database at
a manageable level.
In total, 2\,749\,436 author names are contained in the initial database.

A large fraction of these names are multiple entries of the same author, hence the database 
was post processed to merge identical authors using a sophisticated multi-pass author matching 
scheme that makes use of author names, affiliation identifiers, country codes and activity periods.
A critical parameter of the post processing is the maximum activity gap that is allowed between
a pair of authors for being merged into a single entry.
The rational behind this condition is that it is rather unlikely that an astronomer stops publishing 
articles for a long period, and if there is a long activity gap between a pair of authors, it is likely
that these authors are different individuals.

%%% Figure: Impact of maximum activity gap on the database %%%%%%%%%%%%%%%%%%%
\begin{figure}[!t]
\centering
\includegraphics[width=\columnwidth]{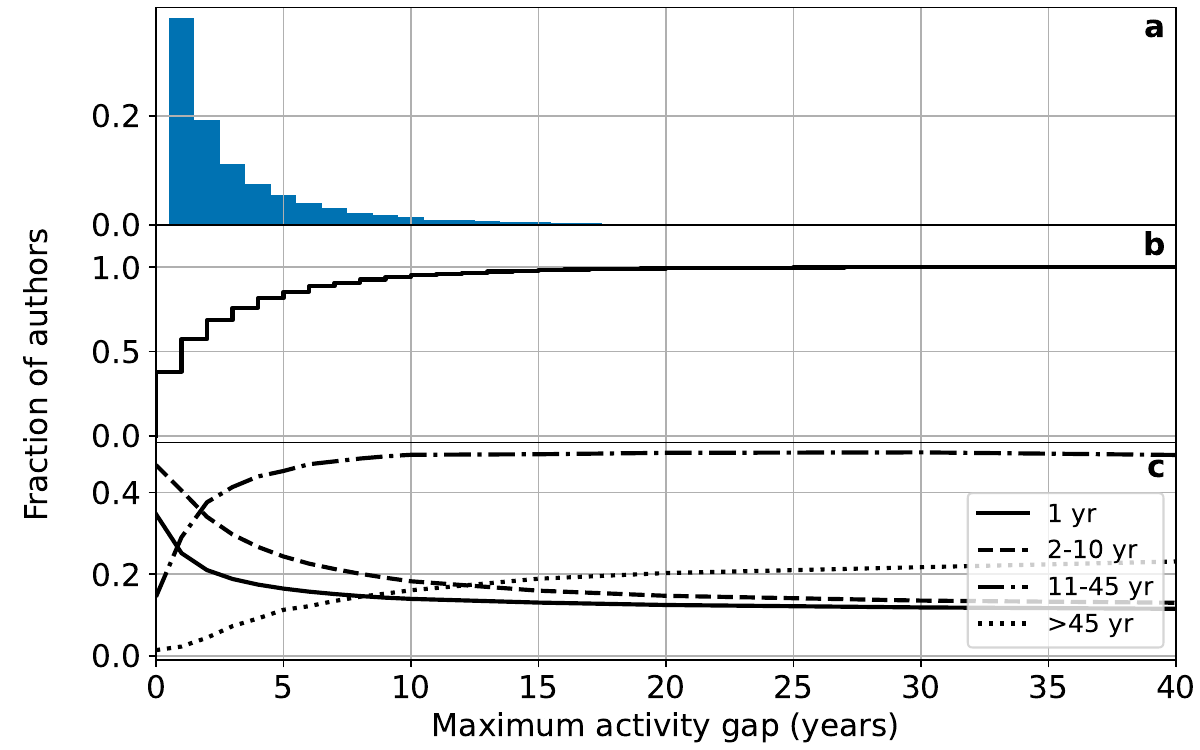}
\caption{
Distribution of maximum activity gaps for a selected set of authors ({\bf a}), its cumulative distribution
({\bf b}) and the fraction of authors in different professional age classes ({\bf c}) as function of the 
maximum activity gap.
\label{fig:maxgap}
}
\end{figure}
%%%%%%%%%%%%%%%%%%%%%%%%%%%%%%%%%%%%%%%%%%%%%%%%%%%

%%% Figure: Impact of maximum activity gap on authors and astronomers %%%%%%%%%%%%%
\begin{figure}[!t]
\centering
\includegraphics[width=\columnwidth]{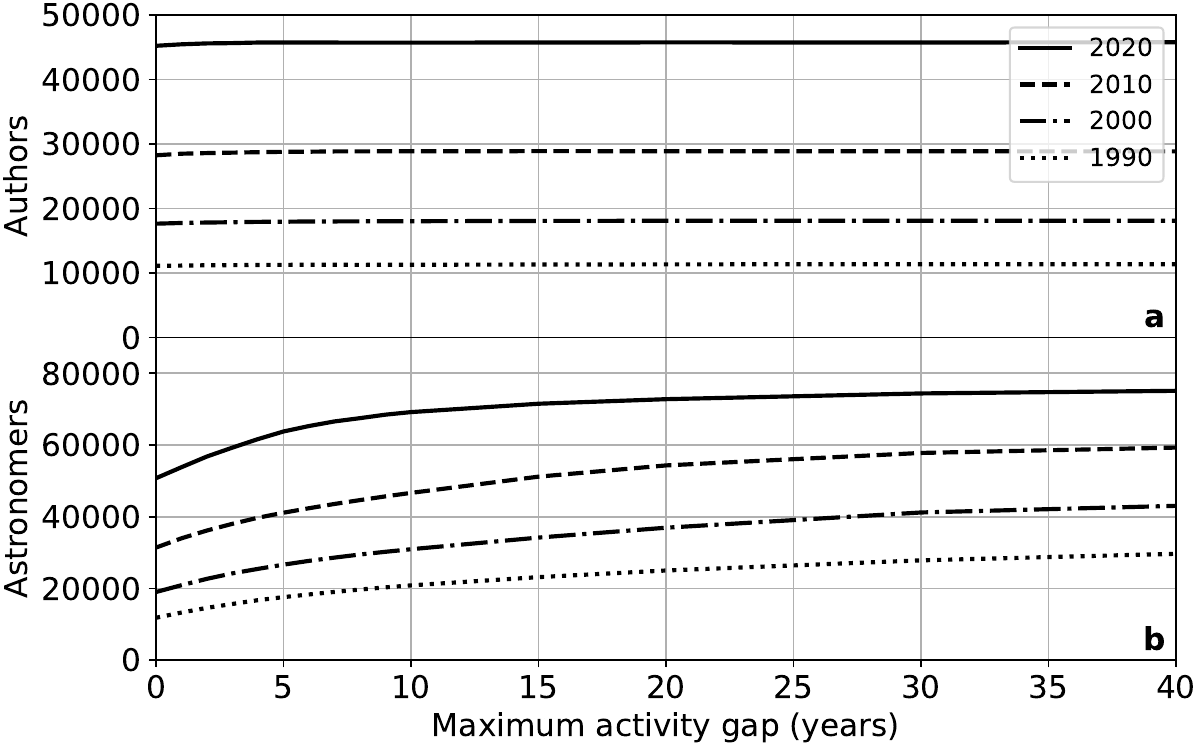}
\caption{
Impact of the choice of the maximum activity gap on the number of authors ({\bf a}) and astronomers
({\bf b}) for the years 1990, 2000, 2010, and 2020.
\label{fig:maxgapimpact}
}
\end{figure}
%%%%%%%%%%%%%%%%%%%%%%%%%%%%%%%%%%%%%%%%%%%%%%%%%%%

To determine a suitable value for the maximum activity gap, the distribution of the longest activity 
gap of the authors was examined.
Since authors with the same name working at a given institution are likely the same person,
authors with fully spelled-out given names and identical affiliation identifiers in the database
were merged and extracted for this purpose.
Panel {\bf a} of Fig.~\ref{fig:maxgap} shows the distribution of the longest activity gaps over
the careers for these authors and panel {\bf b} shows their cumulative distribution.
According to these data, about 38\% of the authors pause publications for one year at most, 
while the remaining authors pause for longer periods, though with a rapidly falling occurrence.
Nevertheless, the distribution has a long tail, and overall, about 80\% of the authors do not pause 
signing publications for more than 5 years, while about 95\% do not pause for more than 10 years.

Panel {\bf c} of Fig.~\ref{fig:maxgap} shows the fraction of authors in different professional age classes,
defined as the mean values of the unshaded curves shown in Fig.~\ref{fig:ageclasses}, as function
of the maximum activity gap parameter.
Recall that age classes are defined by the activity period of an author, with
1 yr being a proxy of PhD students,
2-10 yr a proxy of post-docs,
11-45 yr a proxy of senior scientists, and
$>45$ yr a proxy of retired scientists.
The fraction of PhD students and post-docs decrease with increasing maximum activity gaps, as
longer activity gaps allow for additional mergers that by definition have increased activity periods.
On the other hand, the fraction of retired scientists increases with increasing maximum activity 
gaps, as more and more name fellows are merged that increases the population of scientists with
long activity gaps.
Only the fraction of senior scientists stabilises at a maximum activity gap of about 10 years, which
presents a reasonable compromise between missing only a small number of mergers of the same
individual while avoiding a larger number of mergers of name fellows.

Figure \ref{fig:maxgapimpact} shows the impact of the choice of the maximum activity gap 
parameter on the number of authors (panel {\bf a}) and number of astronomers (panel {\bf b})
for several selected years.
For a given year, the number of authors is basically not affected by the choice of the maximum 
activity gap parameter because it only affects directly the merging of authors that are not 
publishing contemporaneously.
On the other hand, the number of astronomers -- which includes individuals who are not 
publishing in the given year yet that published before and after that year -- increases with
the value of the maximum activity gap parameter.
In other words, the estimate of the number of non-publishing astronomers for a given year is
subject to large uncertainties, making it possibly to provide only a rough estimate of the total
size of the astronomer community.

%%% Figure: Evolution of number of astronomers and authors for selected institutes %%%%%%%%
\begin{figure}
\centering
\includegraphics[width=\columnwidth]{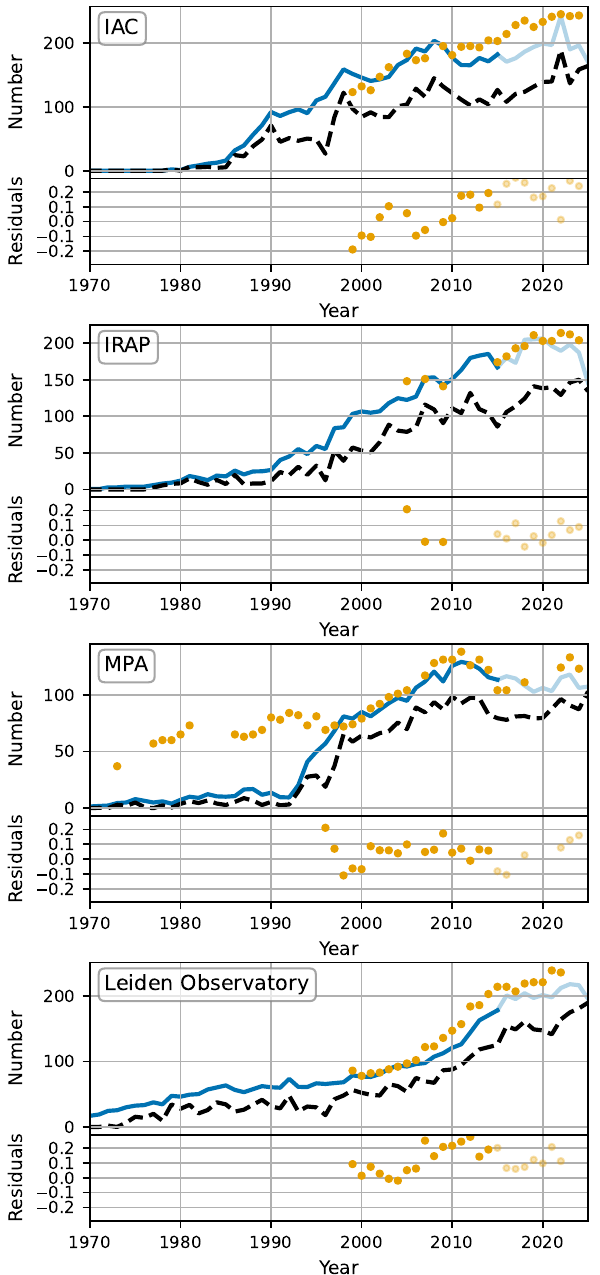}
\caption{
Evolution of the number of authors (black dashed lines) and astronomers (blue solid lines) compared
to the number of scientific staff (orange dots) for a selected number of institutes.
The residuals (scientific staff - astronomers) / astronomers are plotted in the bottom part of each
panel.
Numbers that are affected by the maximum activity gap are shown as shaded lines and symbols.
Scientific staff numbers were gathered from annual institute reports (IAC, Leiden Observatory, MPA)
and from evaluation reports of the Haut Conseil de l'\'evaluation de la recherche et de l'enseignement
sup\'erieur (HCERES) and the former agence d'\'evaluation de la recherche et de l'enseignement
sup\'erieur (AERES) for IRAP.
\label{fig:authors_institutes}
}
\end{figure}
%%%%%%%%%%%%%%%%%%%%%%%%%%%%%%%%%%%%%%%%%%%%%%%%%%%

To assess the amplitude of this uncertainty, Fig.~\ref{fig:authors_institutes} compares the evolution 
of the number of astronomers over the years to the size of the scientific staff for some selected 
institutes.
The institutes were selected on basis of the availability of scientific staff data over extended periods 
in time.
Scientific staff data include tenured personnel, post-docs and graduate students.
A maximum activity gap of 10 years was adopted, and results are shown for
the Instituto de Astrof\'isica de Canarias (IAC) in Spain,
the Institut de Recherche en Astrophysique et Plan\'etologie (IRAP) in France,
the Max Planck Institute for Astrophysics (MPA) in Germany, and
the Leiden Observatory in the Netherlands.
The number of astronomers is roughly within 20\% of the scientific staff numbers.
Note that for MPA, affiliation information is incomplete before the year 1998, which explains why
scientific staff numbers are systematically above the estimate number of astronomers for this
period.

Fixing the value of the maximum activity gap parameter to 10 years, the initial database was
post processed with the goal to merge entries for the same individuals.
The post processing started with a cleaning of author names by removing any characters that 
do not belong to the author name and that probably originate from errors in the ADS article 
ingestion.
The cleaning relies essentially on a visual inspection of the database, resulting in 825 cleaned
author names.

In a second step, author names that do not correspond to individuals but groups of people, such 
as collaborations or teams, were removed from the database.
The removal is based on a list of specific patterns and a list of complete matches that were again
determined by visual inspection of the database.
In total, 3\,388 authors were removed in this step.

In a third step, authors with identical full names and without abbreviated given names were merged 
if either they have the same affiliation or if their maximum activity gap does not exceed 10 years.
99\,951 authors were merged in this step.

%%% Figure: Authors as function of capita, GDP and annual growth rate %%%%%%%%%%%%%%
\begin{figure*}[!t]
\centering
\includegraphics[width=\textwidth]{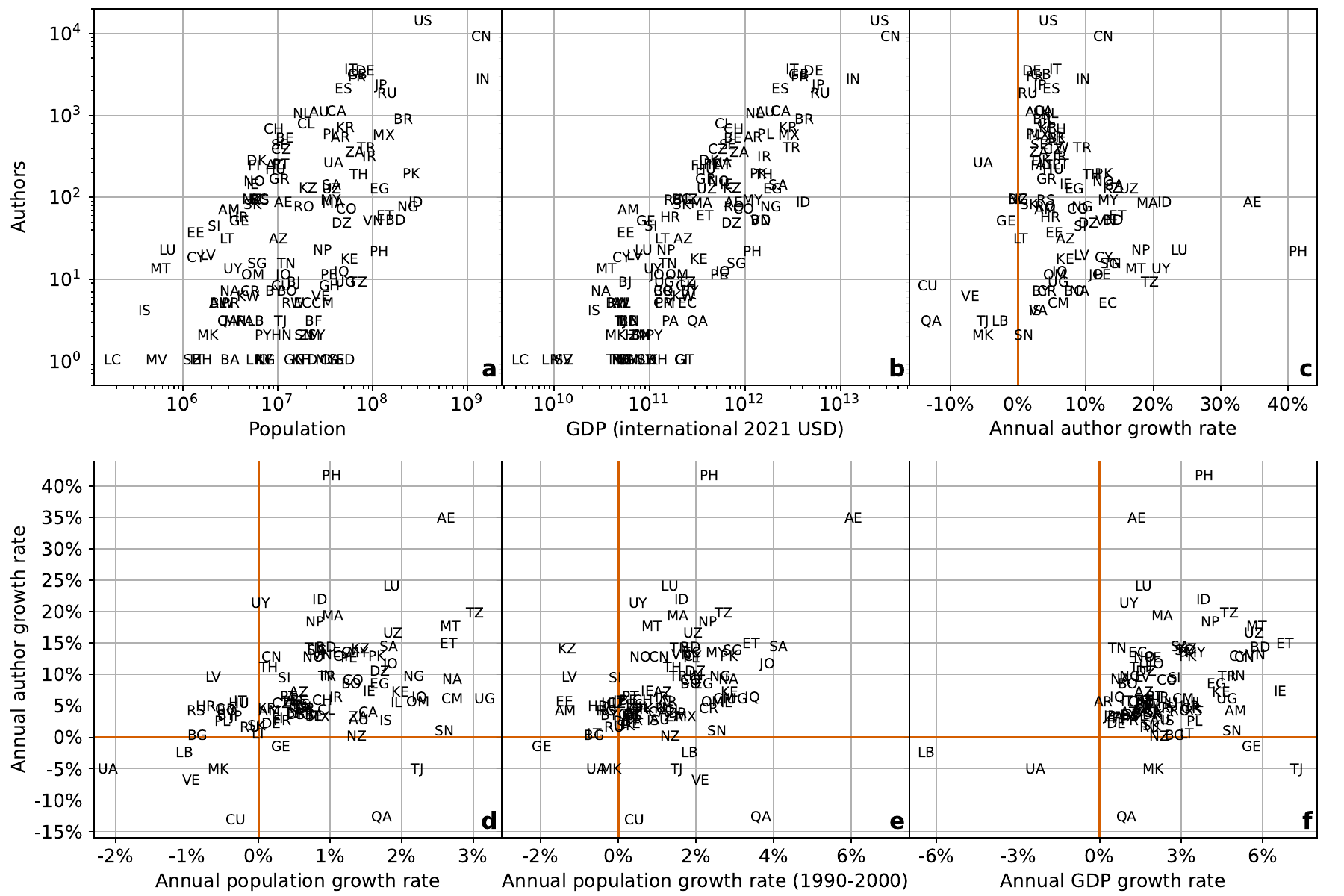}
\caption{
Number of authors in 2025 as function of capita ({\bf a}), GDP ({\bf b}) and annual author growth 
rate ({\bf c}) and annual author growth rate as function of current annual population growth rate ({\bf d}), 
past annual population growth rate ({\bf d}) and annual GDP growth rate ({\bf e}) for all countries. 
Growth rates were determined over the period 2015--2025 except for the past annual population 
growth rate that was determined over the period 1990--2000.
Results for each country are represented by the 2-digit ISO country codes.
\label{fig:authorscapita}
}
\end{figure*}
%%%%%%%%%%%%%%%%%%%%%%%%%%%%%%%%%%%%%%%%%%%%%%%%%%%

In a forth step, authors with identical surnames were analysed for common individuals using their 
given names and the affiliation identifiers.
Fully-spelled given names (such as in ``Public, Joe John'') were considered as potential matches to 
abbreviated given names (such as in ``Public, J. J.''), and all potential matches were scrutinised for 
ambiguities.
Authors were only merged if either no ambiguities exist, or if the ambiguities could be resolved by
using the affiliation identifier.
As merging potentially increases the activity period of the resulting author, the algorithm is applied
five times which is sufficient to catch essentially all possible mergers.
The number of authors merged in the five iterations were 
2\,296\,785, 34\,206, 2\,907, 184 and 8.
This post-processing step led to the largest number of mergers, reducing the number of authors from 
an initial value of  2\,646\,097 at the beginning of this step to 312\,007 at the end.

In a last step, author names were scrutinised for typos by searching for authors with similar names 
using the Ratcliff-Obershelp algorithm \citep{ratcliff1988}.
This algorithm provides for each pair of author names a score between 0 and 1 that quantifies the
match between the author names, with 1 corresponding to identical names.
The algorithm is applied to the full name, as well as only to the surname and given names.
If for a pair of authors the score of all three names is at least equal to 0.8 and if both authors
have the same affiliation, the authors will be merged.
Otherwise, the threshold for merging is raised to 0.9, and typos are only allowed in either the
surname or the given names, meaning that the score for one of those needs to be 1.
Pairs of authors that pass this criterion were merged if ether their host country was identical
or if their host countries and affiliations were either unknown or did not differ.
This reduced the number of authors by 14\,035, resulting in a total number of 297\,972 authors
in the final database.

%%%%%%%%%%%%%%%%%%%%%%%%%%%%%%%%%%%%%%%%%%%%%%%%%%%
% Per country statistics
%%%%%%%%%%%%%%%%%%%%%%%%%%%%%%%%%%%%%%%%%%%%%%%%%%%
\section{Per country community statistics}
\label{appendix:countrydata}

The estimated numbers of authors, authors per capita and authors per GDP are given in 
Table \ref{table:countrydata} for all countries for the year 2025, together with the annual growth
rates over the last decade.
Figure \ref{fig:authorscapita} visualises the numbers and annual growth rates of authors as function 
of potential explanatory variables, such as populations size, GDP, annual growth rates of authors,
annual population growth rate and GDP growth rate.
The figure shows that the number of authors depends little on the size of a country's population,
with variations in the number of authors as large as 2-3 orders of magnitudes for countries with
the same population size.
The dependency of the number of authors on GDP is tighter, yet there is still a substantial
spread of about 1-2 orders of magnitude in the number of authors for countries with the same GDP.
The GDP spread is roughly comparable to that observed by \cite{kurtz2005} and \cite{ahn2019} who 
compared the number of IAU members to GDP.
Both authors suggest that different branches exist in the relation between number of IAU members
and GDP that reflect cultural differences, an idea that is not supported by Fig.~\ref{fig:authorscapita}.
There seems to exist rather a continuity of possible number of authors for a given GDP value that
may be explained by a variety of factors, including besides cultural aspects also economic and
geopolitical interests.

The annual growth rate in the number of authors is largest in countries with a small number of
astronomers, which can be considered as the emerging countries in astronomy.
Among those are 
the Philippines (42\% yr$^{-1}$),
the United Arab Emirates (35\% yr$^{-1}$),
Luxembourg (24\% yr$^{-1}$) and
Indonesia (22\% yr$^{-1}$).
The largest growth rates among the countries with more than 1\,000 authors are found in
China (13\% yr$^{-1}$),
India (10\% yr$^{-1}$),
Italy (6\% yr$^{-1}$)
and Spain (5\% yr$^{-1}$).
There are only eight countries where the number of authors was decreasing over the last
decade.
Among them is Ukraine (-5\% yr$^{-1}$) and Lebanon (-3\% yr$^{-1}$), which are currently under
attack by hostile neighbouring countries, 
Cuba (-13\% yr$^{-1}$),
Venezuela (-7\% yr$^{-1}$)
and Georgia (-2\% yr$^{-1}$)
that are subject to severe economic and political crises,
as well as
North Macedonia (-5\% yr$^{-1}$),
Qatar (-13\% yr$^{-1}$) and 
Tajikistan (-5\% yr$^{-1}$).

The annual growth rate in the number of authors depends little on the growth rates of population
(panel {\bf d}) and GDP (panel {\bf f}).
As one may argue that the population growth rate may only influence the population of 
astronomers after a certain delay, given by the period between birth and PhD, I compare in 
panel {\bf e} also the author growth rate over the period 2015--2025 to the
population growth rate over the period 1990--2000, hence with a delay of 25 years.
Also in this case, no obvious correlation exists between the growth rates of authors and the
population.
Interestingly, while the population was (in 1990--2000) and currently is in decline in some 
countries, the number of authors is still growing in most of those countries.
The only countries where currently both population and number of authors are on decline are
Cuba,
Lebanon,
North Macedonia,
Ukraine and
Venezuela.
GDP is only in decline in two countries (Lebanon, Ukraine) and in both countries also the number
of authors decreases.
Note that no GDP data are provided for Cuba and Venezuela by the world bank, hence both
countries are not shown in panel {\bf f}.
On the other hand, GDP increases in 
Georgia,
North Macedonia,
Qatar and
Tajikistan,
while the number of authors declines.
The case of Qatar and Tajikistan are remarkable because despite the growth of the population,
the number of authors decreases; yet at the same time those countries only have a handful of
authors, hence these tendencies may actually not be very significant.
The Philippines and the United Arab Emirates stand out by a remarkable growth in the number of 
authors that is orders of magnitude larger than the growth in population and in GDP.

%%% Table: 2025 statistics and growth rates %%%%%%%%%%%%%%%%%%%%%%%%%%%%
\begin{longtable*}{l r r r r r r}
\caption{Estimated number of authors for the year 2025. Authors per capita are given per million people,
authors per GDP are given per billion ($10^9$) international 2021 dollars. In each of the three columns,
the first number represents the absolute value while the second number represents the annual growth
rate over the past decade (years 2015--2025), given in percent per year. Growth rates were only 
determined for countries for which the total number of authors other the last decade reached 30.
Author numbers and growth rates for individual countries were corrected for incomplete affiliations by 
dividing author numbers for each year by the fraction of affiliated authors in the world.}
\label{table:countrydata}
\\
\hline\\[0.5mm]
Country & \multicolumn{2}{c}{Authors} & \multicolumn{2}{c}{Authors / capita} & \multicolumn{2}{c}{Authors / GDP}\\[0.5mm]
\hline
\endfirsthead
\caption{continued.}\\
\hline\\[0.5mm]
Country & \multicolumn{2}{c}{Authors} & \multicolumn{2}{c}{Authors / capita} & \multicolumn{2}{c}{Authors / GDP}\\[0.5mm]
\hline
\endhead
\hline
\endfoot
World & 59\,940 & +5.1\% yr$^{-1}$ & 7.30 & +4.2\% yr$^{-1}$ & 0.33 & +2.2\% yr$^{-1}$\\[0.1mm]
Albania & 5 & - & 2.26 & - & 0.10 & -\\[0.1mm]
Algeria & 47 & +10.4\% yr$^{-1}$ & 1.00 & +8.7\% yr$^{-1}$ & 0.06 & +8.7\% yr$^{-1}$\\[0.1mm]
Argentina & 524 & +5.5\% yr$^{-1}$ & 11 & +5.0\% yr$^{-1}$ & 0.41 & +5.4\% yr$^{-1}$\\[0.1mm]
Armenia & 68 & +4.0\% yr$^{-1}$ & 22 & +3.9\% yr$^{-1}$ & 1.05 & -1.0\% yr$^{-1}$\\[0.1mm]
Australia & 1\,078 & +2.6\% yr$^{-1}$ & 39 & +1.2\% yr$^{-1}$ & 0.65 & +0.4\% yr$^{-1}$\\[0.1mm]
Austria & 239 & +3.9\% yr$^{-1}$ & 26 & +3.2\% yr$^{-1}$ & 0.41 & +2.8\% yr$^{-1}$\\[0.1mm]
Azerbaijan & 30 & +7.0\% yr$^{-1}$ & 2.90 & +6.5\% yr$^{-1}$ & 0.13 & +5.4\% yr$^{-1}$\\[0.1mm]
Bahrain & 1 & - & 0.33 & - & 0.005 & -\\[0.1mm]
Bangladesh & 51 & +14.1\% yr$^{-1}$ & 0.29 & +13.2\% yr$^{-1}$ & 0.03 & +8.3\% yr$^{-1}$\\[0.1mm]
Belarus & 7 & +3.4\% yr$^{-1}$ & 0.82 & +3.8\% yr$^{-1}$ & 0.03 & +2.3\% yr$^{-1}$\\[0.1mm]
Belgium & 518 & +5.8\% yr$^{-1}$ & 43 & +5.2\% yr$^{-1}$ & 0.68 & +4.1\% yr$^{-1}$\\[0.1mm]
Benin & 9 & - & 0.61 & - & 0.15 & -\\[0.1mm]
Bolivia & 7 & +8.4\% yr$^{-1}$ & 0.59 & +7.1\% yr$^{-1}$ & 0.05 & +7.0\% yr$^{-1}$\\[0.1mm]
Bosnia and Herz. & 1 & - & 0.34 & - & 0.02 & -\\[0.1mm]
Botswana & 5 & - & 1.86 & - & 0.11 & -\\[0.1mm]
Brazil & 860 & +3.7\% yr$^{-1}$ & 4.04 & +3.2\% yr$^{-1}$ & 0.20 & +2.1\% yr$^{-1}$\\[0.1mm]
Bulgaria & 92 & +0.1\% yr$^{-1}$ & 14 & +0.9\% yr$^{-1}$ & 0.41 & -2.7\% yr$^{-1}$\\[0.1mm]
Burkina Faso & 3 & - & 0.14 & - & 0.05 & -\\[0.1mm]
C\^ote d'Ivoire & 1 & - & 0.03 & - & 0.005 & -\\[0.1mm]
Cambodia & 1 & - & 0.06 & - & 0.008 & -\\[0.1mm]
Cameroon & 5 & +6.0\% yr$^{-1}$ & 0.18 & +3.3\% yr$^{-1}$ & 0.04 & +2.9\% yr$^{-1}$\\[0.1mm]
Canada & 1\,093 & +3.8\% yr$^{-1}$ & 26 & +2.2\% yr$^{-1}$ & 0.45 & +1.8\% yr$^{-1}$\\[0.1mm]
Chad & 1 & - & 0.05 & - & 0.02 & -\\[0.1mm]
Chile & 759 & +4.1\% yr$^{-1}$ & 38 & +3.2\% yr$^{-1}$ & 1.24 & +2.2\% yr$^{-1}$\\[0.1mm]
China & 8\,971 & +12.6\% yr$^{-1}$ & 6.38 & +12.4\% yr$^{-1}$ & 0.25 & +7.3\% yr$^{-1}$\\[0.1mm]
Colombia & 70 & +8.9\% yr$^{-1}$ & 1.30 & +7.6\% yr$^{-1}$ & 0.07 & +6.4\% yr$^{-1}$\\[0.1mm]
Costa Rica & 7 & +4.4\% yr$^{-1}$ & 1.41 & +3.7\% yr$^{-1}$ & 0.05 & +1.0\% yr$^{-1}$\\[0.1mm]
Croatia & 55 & +4.8\% yr$^{-1}$ & 14 & +5.5\% yr$^{-1}$ & 0.32 & +1.5\% yr$^{-1}$\\[0.1mm]
Cuba & 8 & -13.4\% yr$^{-1}$ & 0.78 & -13.0\% yr$^{-1}$ & - & -\\[0.1mm]
Cyprus & 18 & +12.7\% yr$^{-1}$ & 13 & +11.5\% yr$^{-1}$ & 0.33 & +7.6\% yr$^{-1}$\\[0.1mm]
Czechia & 376 & +5.2\% yr$^{-1}$ & 35 & +4.9\% yr$^{-1}$ & 0.70 & +3.6\% yr$^{-1}$\\[0.1mm]
Denmark & 274 & +3.5\% yr$^{-1}$ & 46 & +2.9\% yr$^{-1}$ & 0.63 & +1.5\% yr$^{-1}$\\[0.1mm]
Ecuador & 5 & +13.3\% yr$^{-1}$ & 0.26 & +12.1\% yr$^{-1}$ & 0.02 & +11.9\% yr$^{-1}$\\[0.1mm]
Egypt & 124 & +8.4\% yr$^{-1}$ & 1.05 & +6.7\% yr$^{-1}$ & 0.06 & +4.1\% yr$^{-1}$\\[0.1mm]
Estonia & 36 & +5.4\% yr$^{-1}$ & 26 & +5.0\% yr$^{-1}$ & 0.63 & +3.7\% yr$^{-1}$\\[0.1mm]
Ethiopia & 58 & +14.8\% yr$^{-1}$ & 0.43 & +12.1\% yr$^{-1}$ & 0.14 & +7.9\% yr$^{-1}$\\[0.1mm]
Finland & 237 & +2.8\% yr$^{-1}$ & 42 & +2.6\% yr$^{-1}$ & 0.75 & +2.1\% yr$^{-1}$\\[0.1mm]
France & 2\,862 & +2.5\% yr$^{-1}$ & 42 & +2.2\% yr$^{-1}$ & 0.76 & +1.4\% yr$^{-1}$\\[0.1mm]
Georgia & 50 & -1.7\% yr$^{-1}$ & 13 & -2.0\% yr$^{-1}$ & 0.50 & -7.3\% yr$^{-1}$\\[0.1mm]
Germany & 3\,376 & +2.1\% yr$^{-1}$ & 40 & +1.9\% yr$^{-1}$ & 0.64 & +1.5\% yr$^{-1}$\\[0.1mm]
Ghana & 8 & - & 0.23 & - & 0.03 & -\\[0.1mm]
Greece & 162 & +4.2\% yr$^{-1}$ & 16 & +4.7\% yr$^{-1}$ & 0.41 & +2.6\% yr$^{-1}$\\[0.1mm]
Guatemala & 1 & - & 0.06 & - & 0.004 & -\\[0.1mm]
Guinea & 1 & - & 0.07 & - & 0.02 & -\\[0.1mm]
Honduras & 2 & - & 0.19 & - & 0.03 & -\\[0.1mm]
Hungary & 213 & +5.2\% yr$^{-1}$ & 22 & +5.5\% yr$^{-1}$ & 0.54 & +2.7\% yr$^{-1}$\\[0.1mm]
Iceland & 4 & +2.6\% yr$^{-1}$ & 10 & +0.8\% yr$^{-1}$ & 0.15 & +0.0\% yr$^{-1}$\\[0.1mm]
India & 2\,701 & +9.7\% yr$^{-1}$ & 1.85 & +8.7\% yr$^{-1}$ & 0.18 & +4.6\% yr$^{-1}$\\[0.1mm]
Indonesia & 84 & +21.7\% yr$^{-1}$ & 0.29 & +20.9\% yr$^{-1}$ & 0.02 & +17.9\% yr$^{-1}$\\[0.1mm]
Iran & 305 & +6.2\% yr$^{-1}$ & 3.30 & +5.2\% yr$^{-1}$ & 0.20 & +3.9\% yr$^{-1}$\\[0.1mm]
Iraq & 12 & +6.2\% yr$^{-1}$ & 0.25 & +4.0\% yr$^{-1}$ & 0.02 & +5.6\% yr$^{-1}$\\[0.1mm]
Ireland & 140 & +7.1\% yr$^{-1}$ & 25 & +5.6\% yr$^{-1}$ & 0.19 & +0.5\% yr$^{-1}$\\[0.1mm]
Israel & 241 & +5.3\% yr$^{-1}$ & 24 & +3.4\% yr$^{-1}$ & 0.49 & +1.7\% yr$^{-1}$\\[0.1mm]
Italy & 3\,526 & +5.6\% yr$^{-1}$ & 60 & +5.8\% yr$^{-1}$ & 1.12 & +4.5\% yr$^{-1}$\\[0.1mm]
Japan & 2\,303 & +3.3\% yr$^{-1}$ & 19 & +3.6\% yr$^{-1}$ & 0.39 & +2.9\% yr$^{-1}$\\[0.1mm]
Jordan & 11 & +11.6\% yr$^{-1}$ & 0.97 & +9.9\% yr$^{-1}$ & 0.09 & +9.5\% yr$^{-1}$\\[0.1mm]
Kazakhstan & 126 & +14.0\% yr$^{-1}$ & 6.06 & +12.5\% yr$^{-1}$ & 0.16 & +10.7\% yr$^{-1}$\\[0.1mm]
Kenya & 17 & +7.0\% yr$^{-1}$ & 0.29 & +5.0\% yr$^{-1}$ & 0.05 & +2.5\% yr$^{-1}$\\[0.1mm]
Kuwait & 6 & - & 1.20 & - & 0.02 & -\\[0.1mm]
Kyrgyzstan & 1 & - & 0.10 & - & 0.01 & -\\[0.1mm]
Latvia & 19 & +9.4\% yr$^{-1}$ & 10 & +10.0\% yr$^{-1}$ & 0.26 & +7.8\% yr$^{-1}$\\[0.1mm]
Lebanon & 3 & -2.6\% yr$^{-1}$ & 0.45 & -1.6\% yr$^{-1}$ & - & +6.2\% yr$^{-1}$\\[0.1mm]
Liberia & 1 & - & 0.19 & - & 0.11 & -\\[0.1mm]
Libya & 1 & - & 0.14 & - & 0.01 & -\\[0.1mm]
Lithuania & 30 & +0.4\% yr$^{-1}$ & 10 & +0.4\% yr$^{-1}$ & 0.21 & -2.8\% yr$^{-1}$\\[0.1mm]
Luxembourg & 22 & +23.9\% yr$^{-1}$ & 32 & +22.0\% yr$^{-1}$ & 0.25 & +22.2\% yr$^{-1}$\\[0.1mm]
Madagascar & 1 & - & 0.03 & - & 0.02 & -\\[0.1mm]
Malaysia & 90 & +13.3\% yr$^{-1}$ & 2.50 & +12.0\% yr$^{-1}$ & 0.07 & +9.8\% yr$^{-1}$\\[0.1mm]
Maldives & 1 & - & 2.00 & - & 0.08 & -\\[0.1mm]
Malta & 13 & +17.5\% yr$^{-1}$ & 23 & +14.8\% yr$^{-1}$ & 0.36 & +11.7\% yr$^{-1}$\\[0.1mm]
Mexico & 554 & +3.1\% yr$^{-1}$ & 4.20 & +2.3\% yr$^{-1}$ & 0.19 & +2.2\% yr$^{-1}$\\[0.1mm]
Mongolia & 3 & - & 0.89 & - & 0.05 & -\\[0.1mm]
Morocco & 82 & +19.1\% yr$^{-1}$ & 2.13 & +18.1\% yr$^{-1}$ & 0.22 & +16.8\% yr$^{-1}$\\[0.1mm]
Namibia & 7 & +9.1\% yr$^{-1}$ & 2.23 & +6.4\% yr$^{-1}$ & 0.22 & +8.2\% yr$^{-1}$\\[0.1mm]
Nepal & 22 & +18.2\% yr$^{-1}$ & 0.75 & +17.4\% yr$^{-1}$ & 0.14 & +14.6\% yr$^{-1}$\\[0.1mm]
Netherlands & 1\,030 & +4.6\% yr$^{-1}$ & 57 & +3.9\% yr$^{-1}$ & 0.80 & +2.7\% yr$^{-1}$\\[0.1mm]
New Zealand & 91 & +0.0\% yr$^{-1}$ & 17 & -1.4\% yr$^{-1}$ & 0.35 & -2.2\% yr$^{-1}$\\[0.1mm]
Nicaragua & 1 & - & 0.15 & - & 0.02 & -\\[0.1mm]
Nigeria & 75 & +9.5\% yr$^{-1}$ & 0.32 & +7.4\% yr$^{-1}$ & 0.04 & +8.4\% yr$^{-1}$\\[0.1mm]
North Macedonia & 2 & -5.2\% yr$^{-1}$ & 1.16 & -4.7\% yr$^{-1}$ & 0.05 & -7.2\% yr$^{-1}$\\[0.1mm]
Norway & 151 & +12.7\% yr$^{-1}$ & 27 & +11.9\% yr$^{-1}$ & 0.28 & +11.0\% yr$^{-1}$\\[0.1mm]
Oman & 11 & +5.5\% yr$^{-1}$ & 2.03 & +3.0\% yr$^{-1}$ & 0.06 & +3.9\% yr$^{-1}$\\[0.1mm]
Pakistan & 186 & +12.7\% yr$^{-1}$ & 0.73 & +11.1\% yr$^{-1}$ & 0.13 & +9.5\% yr$^{-1}$\\[0.1mm]
Panama & 3 & - & 0.70 & - & 0.02 & -\\[0.1mm]
Paraguay & 2 & - & 0.30 & - & 0.02 & -\\[0.1mm]
Peru & 11 & +12.5\% yr$^{-1}$ & 0.32 & +11.3\% yr$^{-1}$ & 0.02 & +10.5\% yr$^{-1}$\\[0.1mm]
Philippines & 21 & +41.5\% yr$^{-1}$ & 0.18 & +40.6\% yr$^{-1}$ & 0.02 & +38.0\% yr$^{-1}$\\[0.1mm]
Poland & 571 & +2.4\% yr$^{-1}$ & 16 & +2.9\% yr$^{-1}$ & 0.33 & -1.1\% yr$^{-1}$\\[0.1mm]
Portugal & 248 & +6.3\% yr$^{-1}$ & 23 & +5.9\% yr$^{-1}$ & 0.54 & +4.3\% yr$^{-1}$\\[0.1mm]
Puerto Rico & 5 & - & 1.67 & - & 0.04 & -\\[0.1mm]
Qatar & 3 & -12.9\% yr$^{-1}$ & 1.07 & -14.6\% yr$^{-1}$ & 0.010 & -13.9\% yr$^{-1}$\\[0.1mm]
Romania & 74 & +4.1\% yr$^{-1}$ & 3.90 & +4.5\% yr$^{-1}$ & 0.10 & +1.2\% yr$^{-1}$\\[0.1mm]
Russia & 1\,803 & +1.4\% yr$^{-1}$ & 13 & +1.5\% yr$^{-1}$ & 0.29 & -0.4\% yr$^{-1}$\\[0.1mm]
Rwanda & 5 & - & 0.36 & - & 0.10 & -\\[0.1mm]
Santa Lucia & 1 & - & 5.89 & - & 0.24 & -\\[0.1mm]
Saudi Arabia & 137 & +14.3\% yr$^{-1}$ & 3.70 & +12.4\% yr$^{-1}$ & 0.06 & +11.3\% yr$^{-1}$\\[0.1mm]
Senegal & 2 & +0.9\% yr$^{-1}$ & 0.11 & -1.7\% yr$^{-1}$ & 0.02 & -4.0\% yr$^{-1}$\\[0.1mm]
Serbia & 90 & +4.1\% yr$^{-1}$ & 14 & +5.0\% yr$^{-1}$ & 0.50 & +0.6\% yr$^{-1}$\\[0.1mm]
Singapore & 15 & +13.7\% yr$^{-1}$ & 2.52 & +12.8\% yr$^{-1}$ & 0.02 & +10.5\% yr$^{-1}$\\[0.1mm]
Slovakia & 79 & +1.6\% yr$^{-1}$ & 15 & +1.7\% yr$^{-1}$ & 0.36 & -0.3\% yr$^{-1}$\\[0.1mm]
Slovenia & 43 & +9.3\% yr$^{-1}$ & 20 & +8.9\% yr$^{-1}$ & 0.41 & +6.5\% yr$^{-1}$\\[0.1mm]
South Africa & 342 & +3.1\% yr$^{-1}$ & 5.28 & +1.7\% yr$^{-1}$ & 0.39 & +2.6\% yr$^{-1}$\\[0.1mm]
South Korea & 695 & +4.5\% yr$^{-1}$ & 13 & +4.3\% yr$^{-1}$ & 0.24 & +2.2\% yr$^{-1}$\\[0.1mm]
Spain & 2\,059 & +5.0\% yr$^{-1}$ & 42 & +4.3\% yr$^{-1}$ & 0.85 & +3.3\% yr$^{-1}$\\[0.1mm]
Sudan & 1 & - & 0.02 & - & 0.01 & -\\[0.1mm]
Sweden & 425 & +3.2\% yr$^{-1}$ & 40 & +2.4\% yr$^{-1}$ & 0.63 & +1.7\% yr$^{-1}$\\[0.1mm]
Switzerland & 655 & +5.7\% yr$^{-1}$ & 72 & +4.8\% yr$^{-1}$ & 0.84 & +3.8\% yr$^{-1}$\\[0.1mm]
Syria & 2 & - & 0.08 & - & - & -\\[0.1mm]
Taiwan & 393 & +6.0\% yr$^{-1}$ & - & - & - & -\\[0.1mm]
Tajikistan & 3 & -5.2\% yr$^{-1}$ & 0.29 & -7.4\% yr$^{-1}$ & 0.06 & -12.5\% yr$^{-1}$\\[0.1mm]
Tanzania & 9 & +19.6\% yr$^{-1}$ & 0.13 & +16.6\% yr$^{-1}$ & 0.03 & +15.1\% yr$^{-1}$\\[0.1mm]
Thailand & 183 & +11.0\% yr$^{-1}$ & 2.56 & +10.8\% yr$^{-1}$ & 0.11 & +9.5\% yr$^{-1}$\\[0.1mm]
Trinidad and Tobago & 1 & - & 0.78 & - & 0.02 & -\\[0.1mm]
Tunisia & 15 & +14.0\% yr$^{-1}$ & 1.25 & +13.2\% yr$^{-1}$ & 0.10 & +13.3\% yr$^{-1}$\\[0.1mm]
Turkey & 391 & +9.6\% yr$^{-1}$ & 4.55 & +8.6\% yr$^{-1}$ & 0.12 & +4.9\% yr$^{-1}$\\[0.1mm]
Uganda & 9 & +6.0\% yr$^{-1}$ & 0.17 & +2.8\% yr$^{-1}$ & 0.06 & +1.3\% yr$^{-1}$\\[0.1mm]
Ukraine & 255 & -5.2\% yr$^{-1}$ & 6.55 & -3.1\% yr$^{-1}$ & 0.43 & -2.8\% yr$^{-1}$\\[0.1mm]
United Arab Emirates & 84 & +34.8\% yr$^{-1}$ & 7.30 & +32.2\% yr$^{-1}$ & - & +34.4\% yr$^{-1}$\\[0.1mm]
United Kingdom & 3\,044 & +3.4\% yr$^{-1}$ & 44 & +2.8\% yr$^{-1}$ & 0.81 & +2.3\% yr$^{-1}$\\[0.1mm]
United States of America & 13\,865 & +4.5\% yr$^{-1}$ & 41 & +4.0\% yr$^{-1}$ & 0.53 & +2.1\% yr$^{-1}$\\[0.1mm]
Uruguay & 13 & +21.2\% yr$^{-1}$ & 3.92 & +21.1\% yr$^{-1}$ & 0.12 & +20.1\% yr$^{-1}$\\[0.1mm]
Uzbekistan & 123 & +16.4\% yr$^{-1}$ & 3.32 & +14.6\% yr$^{-1}$ & 0.29 & +10.8\% yr$^{-1}$\\[0.1mm]
Vatican & 4 & +3.0\% yr$^{-1}$ & - & - & - & -\\[0.1mm]
Venezuela & 6 & -7.0\% yr$^{-1}$ & 0.23 & -6.1\% yr$^{-1}$ & - & -\\[0.1mm]
Vietnam & 50 & +13.0\% yr$^{-1}$ & 0.49 & +12.1\% yr$^{-1}$ & 0.03 & +7.3\% yr$^{-1}$\\[0.1mm]
Yemen & 1 & - & 0.03 & - & - & -\\[0.1mm]
Zambia & 2 & - & 0.10 & - & 0.03 & -\\[0.1mm]
eSwatini & 1 & - & 0.84 & - & 0.08 & -\\[0.1mm]
\end{longtable*}
%%%%%%%%%%%%%%%%%%%%%%%%%%%%%%%%%%%%%%%%%%%%%%%%%%%

\end{appendix}

\end{document}